\documentclass[aip,jap,reprint,twocolumn,showpacs,superscriptaddress]{revtex4-1}
\usepackage{amsmath,amssymb,amsfonts,bm,graphicx,dsfont}

\newcommand{\re}{\text{e}}

\newcommand{\ri}{\text{i}}
\newcommand{\rp}{\text{p}}
\newcommand{\rc}{\text{c}}
\newcommand{\rt}{\text{t}}
\newcommand{\rs}{\text{s}}

\newcommand{\rL}{\text{L}}
\newcommand{\rS}{\text{S}}
\newcommand{\rG}{\text{G}}
\newcommand{\rD}{\text{D}}
\newcommand{\rE}{\text{E}}
\newcommand{\rR}{\text{R}}
\newcommand{\rF}{\text{F}}
\newcommand{\rB}{\text{B}}
\newcommand{\Tr}{\text{Tr}}

\newcommand{\kB}{k_\text{B}}
\newcommand{\kb}{k_\text{B}}

\renewcommand{\Re}{{\rm Re}}
\renewcommand{\Im}{{\rm Im}}

\begin{document}

\title{Contactless heat flux control with photonic devices}

\author{Philippe Ben-Abdallah}
\email{pba@institutoptique.fr}
\affiliation{Laboratoire Charles Fabry, UMR 8501, Institut d'Optique, CNRS, Universit\'e Paris-Sud 11, 2, Avenue Augustin Fresnel,
91127 Palaiseau Cedex, France}
\author{Svend-Age Biehs}
\email{s.age.biehs@uni-oldenburg.de}
\affiliation{Institut f\"ur Physik, Carl von Ossietzky Universit\"at, D-26111 Oldenburg, Germany}


\begin{abstract}
The ability to control  electric currents in solids using diodes and transistors is 
undoubtedly at the origin of the main developments in modern electronics which have 
revolutionized the daily life in the second half of 20th century. Surprisingly, until 
the year 2000 no thermal counterpart for such a control had been proposed. Since then, 
based on pioneering works on the control of phononic heat currents new devices were proposed
which allow for the control of heat fluxes carried by photons rather than phonons or electrons. 
The goal of the present paper is to summarize the main advances achieved recently in the 
field of thermal energy control with photons.
\end{abstract}


\maketitle

%
%

\section{Introduction}
The diode and the transistor introduced by F. Braun~\cite{Braun} and Bardeen {\itshape et al.}~\cite{Bardeen} (based on the works
of Julius E. Lilienfeld from 1925), in 1874 and 1948 are the main building blocks of almost all modern systems of information 
treatment. These elementary devices allow for rectifying, switching, modulating and even 
amplifying the electric current in solids. Similar devices which would provide the same 
degree of control of heat currents instead of the control of electric currents are not 
as widespread in our daily life and the concepts 
for such devices for heat flow management at the nanoscale were introduced as recently 
as about ten years ago. In 2006 Baowen Li {\itshape et al.}~\cite{Casati1} have proposed 
a thermal counterpart of a field-effect transistor. In this device the temperature bias 
plays the role of the voltage bias and the heat currents carried by phonons play the role 
of the electric currents. Later, several prototypes of phononic thermal logic gates~\cite{BaowenLi2} 
as well as thermal memories~\cite{BaowenLi3,BaowenLiEtAl2012} have been developed in order to 
process information by phononic heat currents rather than by electric currents. 

However, this technology suffers from some weaknesses of fundamental nature which intrinsically limit its 
performance. One of the main limitations comes probably from the speed of acoustic phonons (heat carriers) 
which is four or five orders of magnitude smaller than the speed of light and therefore also 
some orders of magnitude smaller than the speed of electrons. Another intrinsic limitation of 
phononic devices is related to the inevitable presence of local Kapitza resistances. These resistances 
which originate from the mismatch of vibrational modes at the interface of different elements 
can reduce the phononic heat flow dramatically. Finally, the strong nonlinear phonon-phonon 
interaction mechanism makes the phononic devices difficult to deal with in presence of a 
strong thermal gradient. On the contrary the physics of energy transport mediated by photons 
instead of phonons remains unchanged close and far from thermal equilibrium. 
These limitations and difficulties explain, in part, why so many efforts have been deployed, 
during the last decades, to develop full optical or at least opto-electronic architectures for 
processing and managing information. Particularly important developments have been carried 
out during the last decade with plasmonics systems~\cite{Ozbay,Caglayan} with the goal to 
increase the speed of information processing substantially while reducing the dimension 
of devices to the nanoscale at the same time. 

In this paper we review the recent developments made to achieve a control of heat flow with 
contactless devices using thermal photons. We discuss the operating principles of the three 
basic building blocks for heat flow management which are the photonic thermal diode~\cite{PBA_APL}, 
photonic thermal transistor~\cite{PBA_PRL2014} and photonic thermal memory~\cite{Slava} based 
on phase-change materials. 

%
%

\section{Radiative thermal diode}

\begin{figure}
\includegraphics[scale=0.3]{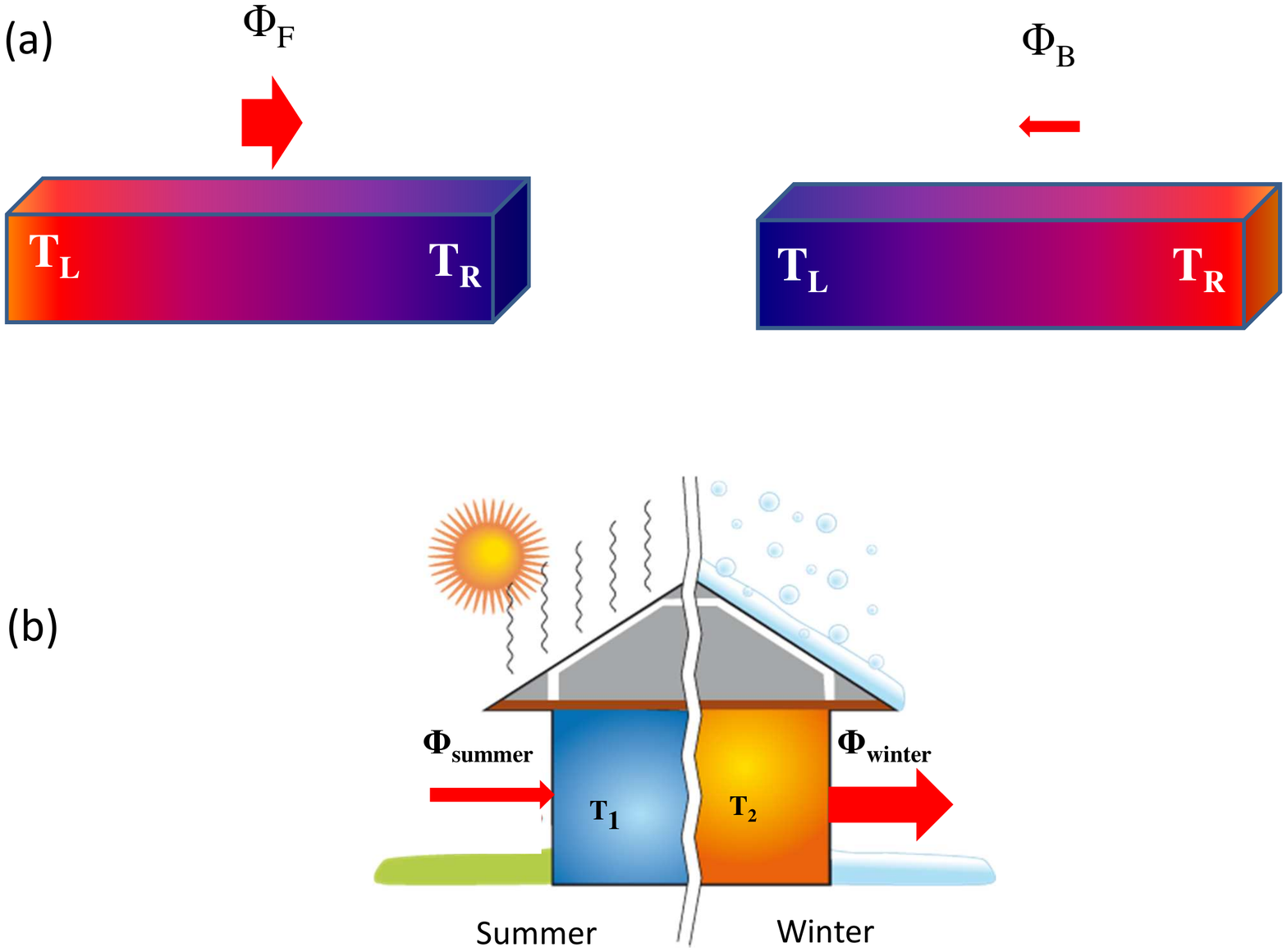} 
\caption{(a) Sketch of thermal rectification principle.  Forward scenario (top-left) : the left side 
         of a medium is in contact with a hot reservoir at temperature $T_\rL$ while its right side is in contact 
         with a cold reservoir at temperature $T_\rR<T_\rL$ so that a heat flux $\Phi_\rF$ is exchanged between  both 
         reservoirs through the medium. Backward scenario (top-right) : the temperature gradient is reversed resulting in 
         a heat flux $\Phi_\rB$ in the opposite direction. Due to a strongly nonlinear thermal behavior of the medium 
         the magnitude of flux $\Phi_\rB$ and for $\Phi_F$ can be different resulting in a rectification of the heat flux
         in one preferred direction. (b) Application of radiative heat flux rectification to a passive thermal 
         regulation of a building.  The non-linear behavior of insulating materials allows to reduce both 
         heating/cooling during summer and winter.}
\label{rectification}
\end{figure}

Asymmetry of heat transport with respect to the sign of the temperature gradient between two 
points Fig.~\ref{rectification}(a) is the basic definition of thermal rectification~\cite{Starr1936,RobertsWalker2011} 
which is at the heart of a variety of applications as, for example, in thermal regulation 
of buildings as scetched in Fig.~\ref{rectification}(b). Usually, this dissymmetry in 
the thermal behaviour of a system finds its origin in the dependence of the 
electical/phononic/optical properties of materials with respect to the temperature. The effectiveness 
of the thermal rectification is generally measured by means of the normalized rectification 
coefficient which can be defined as
\begin{equation}
  \eta=\frac{\mid \Phi_F-\Phi_B\mid}{\text{max}( \Phi_F, \Phi_B)}
\end{equation}
where  $\Phi_F$  and  $\Phi_B$ denote the heat flux in the forward and backward operating 
mode, respectively. Different solid-state thermal rectifiers have been suggested during 
the last decade based on nonlinear atomic vibrations~\cite{BaowenLi2004,Chang}, a nonlinear
dispersion relation of the electron gas in metals~\cite{Segal}, direction dependent Kapitza 
resistances~\cite{Cao} or based on the dependence of the superconducting density of states and phase 
dependence of heat currents ﬂowing through Josephson junctions~\cite{Martinez}.

\begin{figure}
\includegraphics[scale=0.30]{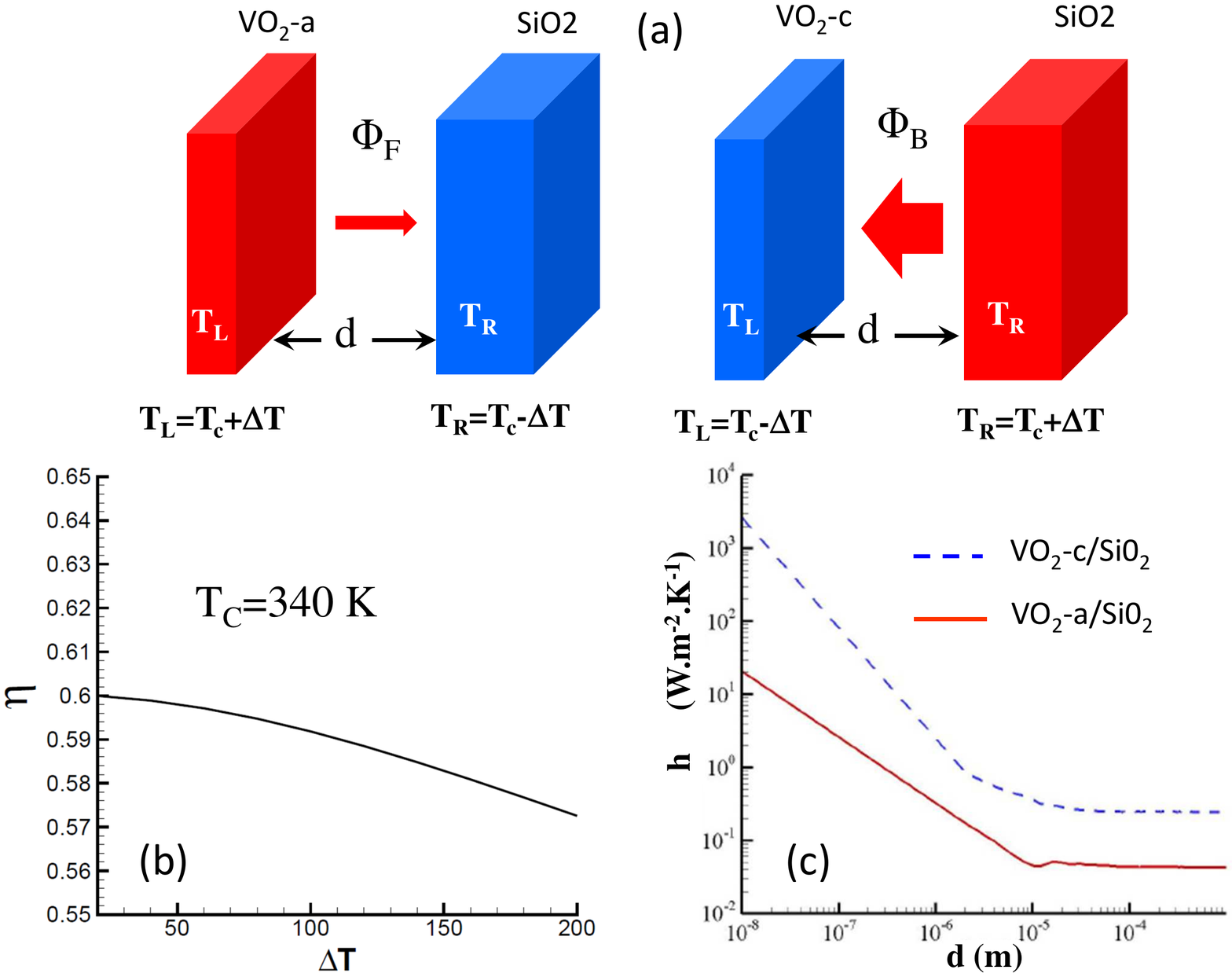} 
\caption{(a) Phase-change radiative thermal diode.  In the forward scenario (top-left) a IMT 
         phase-change material (here VO$_2$) is in its metallic state (VO$_2$-a) at a temperature higher than $T_\rc$.
         In the backward scenario (top right), the phase-change material is in its insulating state (VO$_2$-c). 
         (b) Rectification coefficient in far-field regime of a VO$_2$/SiO$_2$ system with respect to the 
         temperature difference $\Delta T$. The phase transition of VO$_2$ occurs at $T_\rc=300\,{\rm K}$. 
         (c) Heat transfer coefficent $h$ evaluated at $T = T_\rc$ for VO$_2$ is in its metallic or insulating 
          state, respectively. $h$ is plotted with respect to the separation distance $d$.}.
\label{diode}
\end{figure}

During the last five years, photon-mediated thermal rectifiers
have been proposed  both in planar~\cite{OteyEtAl2010,Iizuka,Fan,BasuFrancoeur2011,NefzaouiEtAl2013,Zhang2,Huang,Dames} and non-planar geometry~\cite{Zhu2} by several authors to tune radiative heat exchanges both in near field (i.e. for separation 
distances smaller than the thermal wavelength) and far field (i.e. for separation distances larger than 
the thermal wavelength) using materials with temperature dependent optical and scattering properties. However, in plan geometry, only 
relatively weak thermal rectifications coefficients were achieved with the proposed mechanisms ($\eta< 44\%$ in 
Refs.~\cite{OteyEtAl2010,Iizuka}, $\eta < 52\%$ in Ref.~\cite{BasuFrancoeur2011}). Unfortunately, these 
rectification coefficients are not sufficiently high to operate such devices as radiative diodes.
Very recently another rectification mechanism driven by the phase-change of material properties 
was proposed in order to increase significantly the rectification coefficient both in 
near-field~\cite{van Zwol1}  and far-field regime~\cite{PBA_APL}. In such insulator-metal 
transition (IMT) materials, a small change of the temperature around its critical 
temperature $T_\rc$ causes a sudden qualitative and quantitative change of the material 
properties from an insulating to a metallic behaviour by a Mott transition~\cite{Mott}. Thanks to the 
IMT the optical properties~\cite{Baker} of such materials undergo a rapid change as well resulting 
in thermal rectification coefficients as large as $60\%$ in far-field regime (Fig.~\ref{diode}-b) and 
$99.9\%$ in near-field regime (Fig.~\ref{diode}-c).

To illustrate this, let us consider a system as depicted in Fig.~\ref{diode}, where two plane 
samples, one made of VO$_2$ and one made of amorphous glass (SiO$_2$) at temperatures $T_L$ and $T_R$, 
respectively. Both media are separated by a vacuum gap of thickness $d$.
Here, VO$_2$ is a IMT material which has a critical temperature of
about $T_\rc=340\, {\rm K}$. For temperatures below $T_\rc$  VO$_2$ is in an insulating
phase and has a monocilinic structure and for temperatures above $T_\rc$ it is in a metallic 
phase with a rutile structure. The optical properties~\cite{Baker} can be modelled by a 
uni-axial permittivity for $T < T_\rc$ and by a isotropic permittivity for $T > T_\rc$.

We now examine this system in the two following thermal operating modes: 
(i) In the forward mode (F) the temperature $T_\rL = T_\rc + \Delta T$ of the VO$_2$
slab is greater than its critical temperature $T_\rc$  so that VO$_2$ is in its 
metallic phase while the temperature of the glass medium is set to $T_\rR=T_c-\Delta T < T_\rL$. 
Obviously the average temperature is then centered around $T_\rc$ and the
net radiative heat flow is from the left-hand side to the right.  
(ii) In the backward mode (B) we choose $T_\rL = T_\rc-\Delta T$ so that VO$_2$ 
is in its insulating phase and $T_\rR = T_\rc+\Delta T > T_\rL$. Again the
average temperature is centered around $T_\rc$ but the net energy flow is this time
from the right-hand side to the left.

The exchanged radiative heat flux per unit surface between two semi-infinite media 
can be written in the general form~\cite{Polder1971,BiehsEtAl2010,JoulainPBA2010}
\begin{equation}
\begin{split}
  \Phi_{\rF/\rB} &= \int_0^\infty\!\frac{{\rm d}\omega}{2 \pi} \Delta\Theta(\omega)\!\! \sum_{j = \{\rm s,p\}}\int\! \frac{{\rm d}^2 \boldsymbol{\kappa}}{(2 \pi)^2} \, \mathcal{T}_{j,\rF/\rB}(\omega,\boldsymbol{\kappa}; d) \\
             &= \int_0^\infty\!{\rm d}\omega\, \Delta\Theta(\omega) \varphi_{\rF/\rB}(\omega,d) 
\end{split}
\label{Eq:SpectralPoynting}
\end{equation}
where $\Delta\Theta(\omega)=\Theta(\omega,T_\rL)-\Theta(\omega,T_\rR)$ is the difference of mean energies of 
Planck oscillators at frequency $\omega$ 
\begin{equation}
  \Theta(\omega,T) = \frac{1}{\exp\bigl({\frac{\hbar \omega}{\kB T_{i/j}}}\bigr)-1)}
\end{equation}
at the temperatures of two interacting media; $\kB$ is Boltzmann's constant and $2 \pi \hbar$ is Planck's constant. As 
for $\mathcal{T}_{j,\rF/\rR}(\omega,\boldsymbol\kappa)$ represents the energy transmission coefficient of each 
mode $(\omega,\boldsymbol{\kappa})$ ($\boldsymbol{\kappa}$ is the wave vector parallel to the interfaces) 
for the two polarization states (s and p polarization). For in general anisotropic media it is defined 
for the propagating modes ($\kappa < \omega/c$) by~\cite{Bimonte2009,BiehsEtAl2011}
\begin{equation} 
   \mathcal{T}_{j,\rF/\rB} =
    \Tr\bigl[(\mathds{1} - \mathds{R}_\rR^\dagger \mathds{R}_\rR)  \mathds{D}_{\rL\rR}(\mathds{1} - \mathds{R}_\rL^\dagger \mathds{R}_\rL)  {\mathds{D}_{\rL\rR}}^\dagger \bigr]
\label{Eq:TransmissionCoeff}
\end{equation}
and for the evanescent modes ($\kappa > \omega/c$) by
\begin{equation}
   \mathcal{T}_{j,\rF/\rB} =
    \Tr\bigl[(\mathds{R}_\rR^\dagger - \mathds{R}_\rR) \mathds{D}_{\rL\rR} (\mathds{R}_\rL - \mathds{R}_\rL^\dagger)  {\mathds{D}_{\rL\rR}}^\dagger \bigr]\re^{-2 |k_{z0}| d}.
\label{Eq:TransmissionCoeff2}
\end{equation}
introducing the component of the wave vector in the vacuum gap normal to the interfaces $k_{z0}^2 = \frac{\omega^2}{c^2} - \kappa^2$.
Note that the difference between $\mathcal{T}_{j,\rF}$ and $\mathcal{T}_{j,\rB}$ is caused by the choice of the temperatures
for the two modi (i) and (ii). The reflection matrices of each interface depend on these temperatures through the optical properties
of the media. They are given by ($l = \rL,\rR$)
\begin{eqnarray}
\label{ReflectionMatrices}
\mathds{R}_l = \left[
\begin{array}{cc}
   r^{{\rm s, s}}_l (\omega, \kappa) &  r^{{\rm s, p}}_l (\omega, \kappa) \\
   r^{{\rm p, s}}_l (\omega, \kappa) &  r^{{\rm p, p}}_l (\omega, \kappa) 
\end{array} \right] .
\end{eqnarray}
The matrix $\mathds{D}^{\rL\rR}$ is defined as
\begin{equation}
 \mathds{D}^{\rL\rR}= {(\mathds{1} - \mathds{R}_\rL\mathds{R}_\rR \re^{2 \ri k_{z0} d})}^{-1}.
\end{equation}
The matrix elements $r^{j,j'}_l$ of the reflection matrix are the reflection coefficients for the scattering of an 
incoming $j$-polarized plane wave into an outgoing $j'$-polarized wave. For isotropic or uniaxial media where 
the optical axis is orthogonal to the surface there is no depolarisation, i.e.\ we have $ r^{{\rm s, p}}_l = r^{{\rm p, s}}_l=0$. 
The remaining reflection coefficients are then given by
\begin{equation}
  r^{{\rm s, s}}_l=\frac{k_{z0}-k_{l;s}}{k_{z0}+k_{l;s}},
  \label{Eq:r_ss}
\end{equation}
\begin{equation}
  r^{{\rm p, p}}_l=\frac{\epsilon_{l,\parallel} k_{z0}-k_{l;p}}{\epsilon_{l,\parallel} k_{z0}+k_{l;p}},
  \label{Eq:r_pp}
\end{equation}
where $k_{l;s,p}$ are solutions of the Fresnel equation~\cite{YehBook}
\begin{equation}
  \biggl(\epsilon_{l,\parallel} \frac{\omega^2}{c^2}-\kappa^2-k_{l;s}^2\biggl)\biggr(\epsilon_{l,\parallel} \epsilon_{l, \perp}\frac{\omega^2}{c^2}-\epsilon_{l, \parallel} \kappa^2-\epsilon_{l,\perp}k_{l;p}^2 \biggr) = 0.
\label{Eq:Fresnel}
\end{equation}
Here $\epsilon_{l,\parallel}$ and $\epsilon_{l,\perp}$ are the permittivities parallel and perpendicular to the surface of
the uniaxial material VO$_2$ assuming that the optical axis of VO$_2$ is normal to the surface. 
For amorphous glass which is isotropic we have $\epsilon_\parallel = \epsilon_\perp = \epsilon_{\text{SiO}_2}$~\cite{Palik}. 

With the above expressions we can determine the heat flux in the forward and backward case for different temperature differences $\Delta T$.
The resulting rectification coefficients $\eta$ of the system in Fig.~\ref{diode}(a) are plotted in Figure~\ref{diode}(b) in far-field regime. 
It decreases monotonously with the temperature difference $\Delta T$ between both samples. For very small temperature differences we see 
that $\eta\approx60\%$ illustrating the high rectification efficiency of the phase transition of VO$_2$. This efficiency 
is very large given the fact that we use the simplest possible geometrie. This prediction has been recently experimentally verified~\cite{Ito}. This rectification process could  probably be futher improved by texturing the medium as suggested in Ref.~\cite{NefzaouiEtAl2013}. 

\begin{figure}
\includegraphics[scale=0.26]{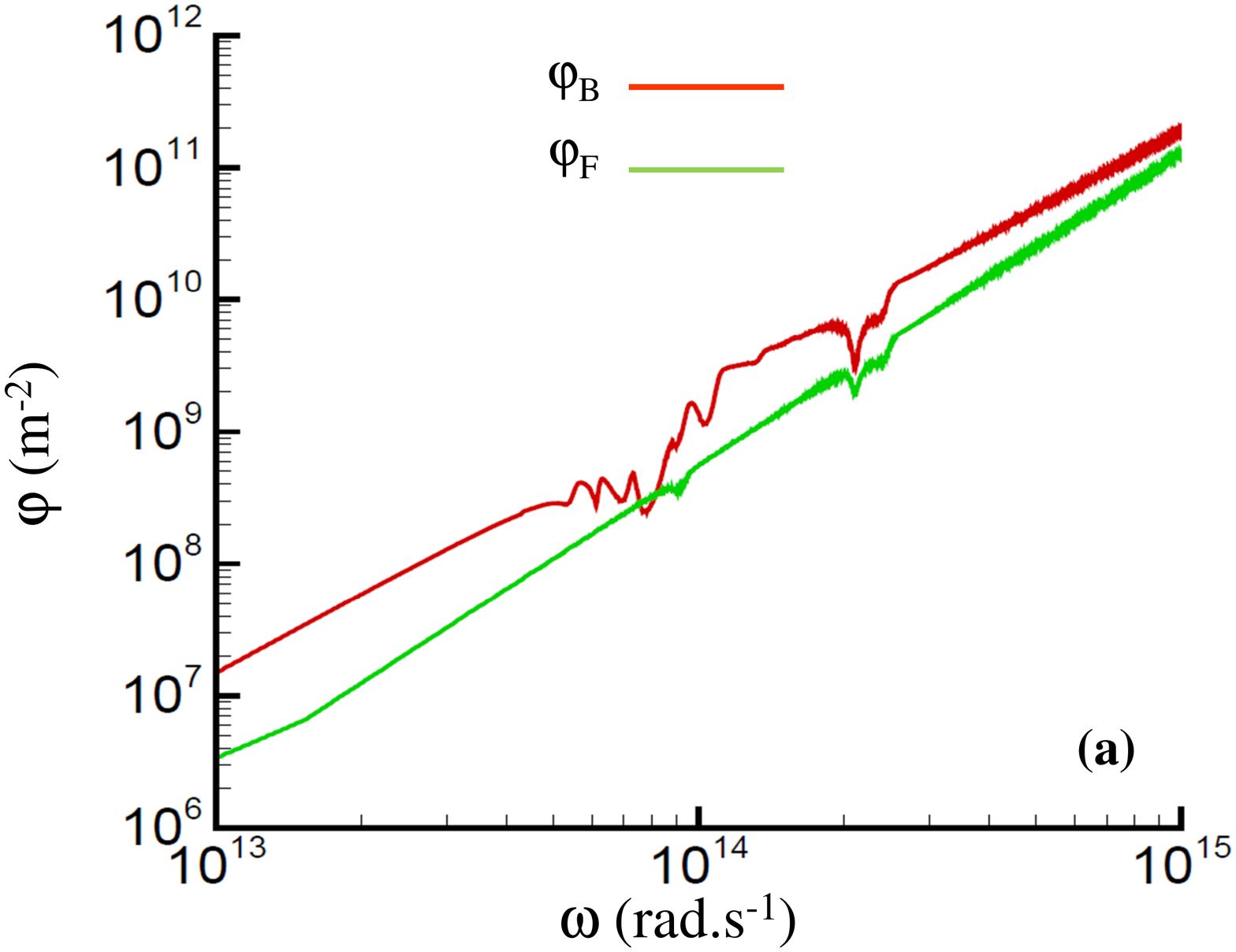} 
\includegraphics[scale=0.25]{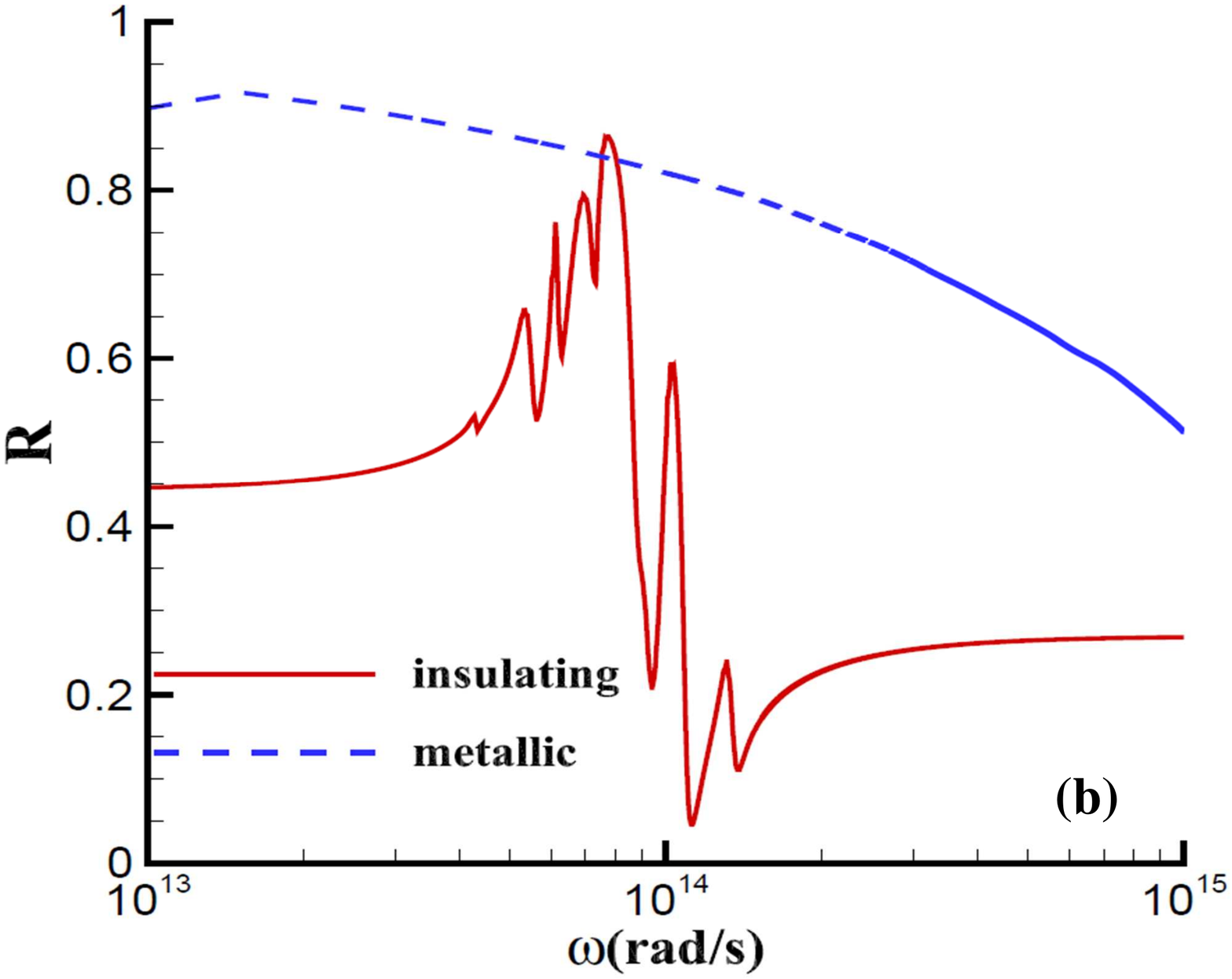} 
\caption{(a) Plot of the spectral heat flux $\varphi_{\rF/\rB}$ (in far-field regime) introduced in Eq.~(\ref{Eq:SpectralPoynting}) 
         as a function of frequency for forward direction where VO$_2$ is in its metallic phase
         and the reverse situation where VO$_2$ is in its insulating phase. In addition, we have plotted 
         $\varphi = \frac{\omega^2}{\pi^2 c^3} \frac{c}{4}$ for the case that both materials are perfect black 
         bodies, i.e.\ $\mathcal{T}_{j,\rR/\rB} = 1$ for all propagating modes.
         (b) Plot of the reflectivity $R$ of VO$_2$ in its insulating and metallic phase for normal incidence as function of frequency.}
\label{spectrum}
\end{figure}

In order to understand this large difference of the radiative heat flow in forward and backward direction we have plotted 
in Fig.~\ref{spectrum} the spectral heat flux $\varphi(\omega,d\rightarrow \infty)$ defined in Eq.~(\ref{Eq:SpectralPoynting}) 
using the transmission coefficient (\ref{Eq:TransmissionCoeff}) for the forward and backward scenario. 
We note that when VO$_2$ is in its metallic phase (forward scenario, $T_\rL>T_c$), the spectral heat flux is broadband and scales 
approximately like $\propto \omega^2$. The same is true when VO$_2$ is in its insulating phase. The difference is
that in the insulating phase the spectral heat flux is generally larger and shows much more structure. This can be understood
by the reflectivity of VO$_2$ which is plotted in  Fig.~\ref{spectrum}(b). In the metallic phase the reflectivity of VO$_2$ is
relatively high as can be expected from a metallic material whereas in the insulating phase the reflectivity is generally much
smaller in the shown frequency range. Therefore it follows directly from Kirchhoff's law that the emissivity of VO$_2$ in the
metallic phase is smaller than in the insulating phase. This asymetry is the key for obtaining a highly efficient 
phase-change thermal diode. It is interesting to note, that in the spectral region around $10^{14}\,{\rm rad}/{\rm s}$ the reflectivity
of VO$_2$ in its insulating phase can also be relatively large and attain values close or even larger than that in its
metallic phase. In this frequency band, the insulating VO$_2$ is metal-like which means here that the permittivity can be 
negative. In this regime VO$_2$ in its insulating phase supports surface polaritons.

\begin{figure}
\includegraphics[scale=0.35,angle=90]{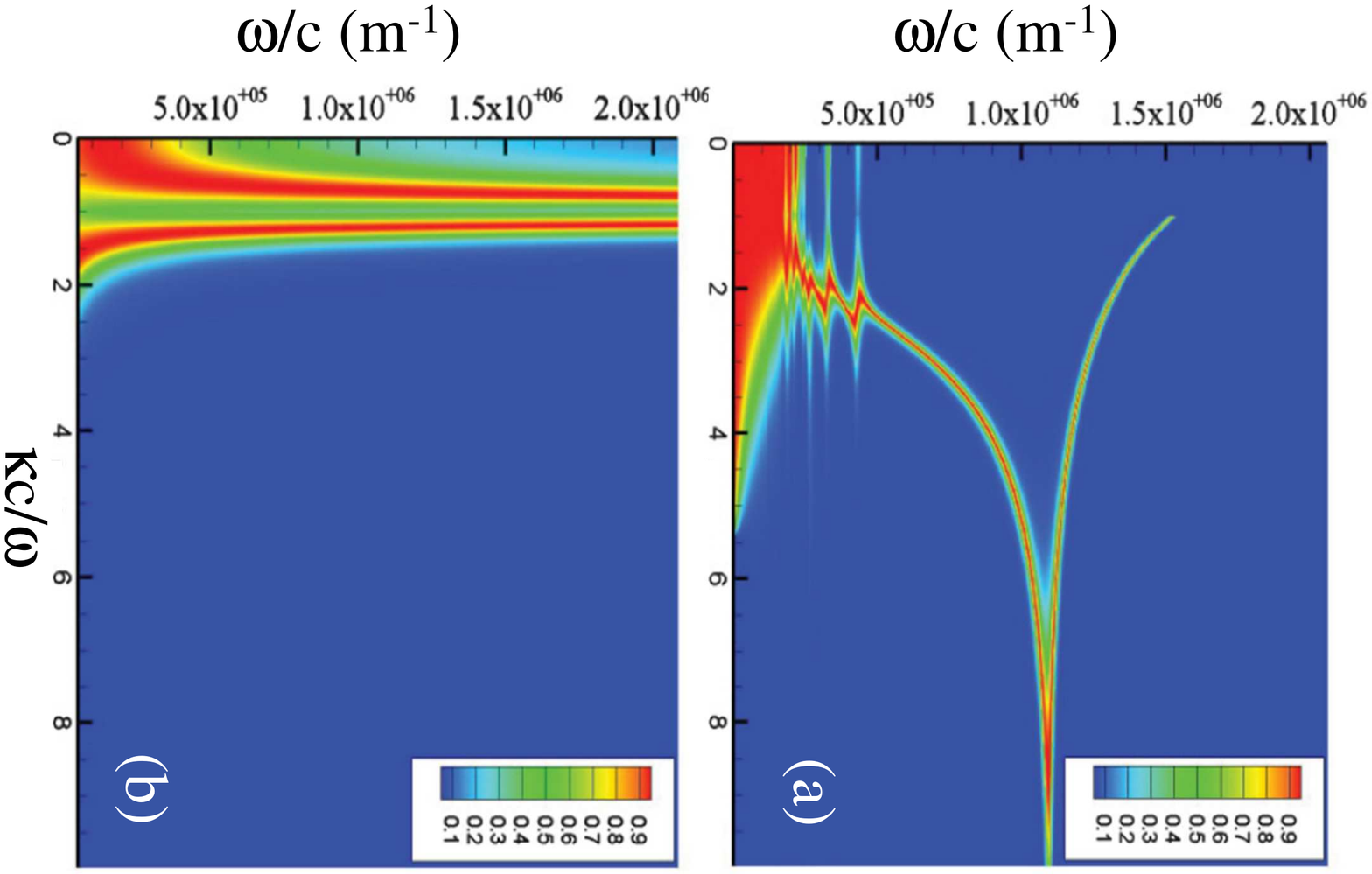} 
\caption{(a) Transmission coefficient $\mathcal{T}_{\rp,\rR/\rB}$ (p polarization)
         between two insulating VO$_2$ surfaces and (b) between two metallic VO$_2$ 
         surfaces. Both graphs were obtained for a separation distance $d = 500\,{\rm nm}$. 
         The spectral character of thermal transfer changes radically across 
         the phase transition}.
\label{trans_eva}
\end{figure}

In near-field regime that is to say when the separation distance between the VO$_2$ and SiO$_2$ samples is smaller 
than the thermal wavelength, the 'diodicity' of system becomes even better as we see on Figure~\ref{diode}(c) by 
comparing the heat transfer coefficients
\begin{equation}
  h_{\rF/\rB} = \frac{\partial \Phi_{\rF/\rB}}{\partial T}\biggr|_{T = T_\rc}
\end{equation}
for the forward and backward scenarios as a function of the separation distance $d$ when temperature is equal to $T_c$.  
We observe a very large thermal contrast of about two orders of magnitude in the near field against about a 
factor 5 in the far field. In order to understand the physics involved in this spectacular increasement, we have 
plotted in Fig.~\ref{trans_eva} the the transmission coefficients for both phases of VO$_2$ in the ($\omega,k$) 
plane at a given separation distance in near-field regime. For insulating VO$_2$ we see in Fig.~\ref{trans_eva}(a) 
that this coefficient is close to one far beyond the light line $\kappa =\omega/c$. In this region the contribution
of the surface phonon polaritons of VO$_2$ which couple to the surface polaritons of SiO$_2$ can be observed 
(see \cite{van Zwol1} for a detailed discussion). As the coupling is rather efficient in this region, the heat 
transfer is large. On the other hand, when VO$_2$ is metallic we obtain the transmission coefficient shown in 
Fig.~\ref{trans_eva}(b), the surface phonon-polariton contribution disappears and as a consequence the heat 
transfer is small. A detailed discussion of the efficiency of rectification in such diode as a function of 
the thickness of the slabs can be found in Ref.~\cite{LipingWang}. Also, based on a similar idea, a contactless thermal switch has been recently suggested~\cite{Switch}. 

%
%

\section{Photonic thermal transistor}

\begin{figure}
\includegraphics[scale=0.35,angle=-90]{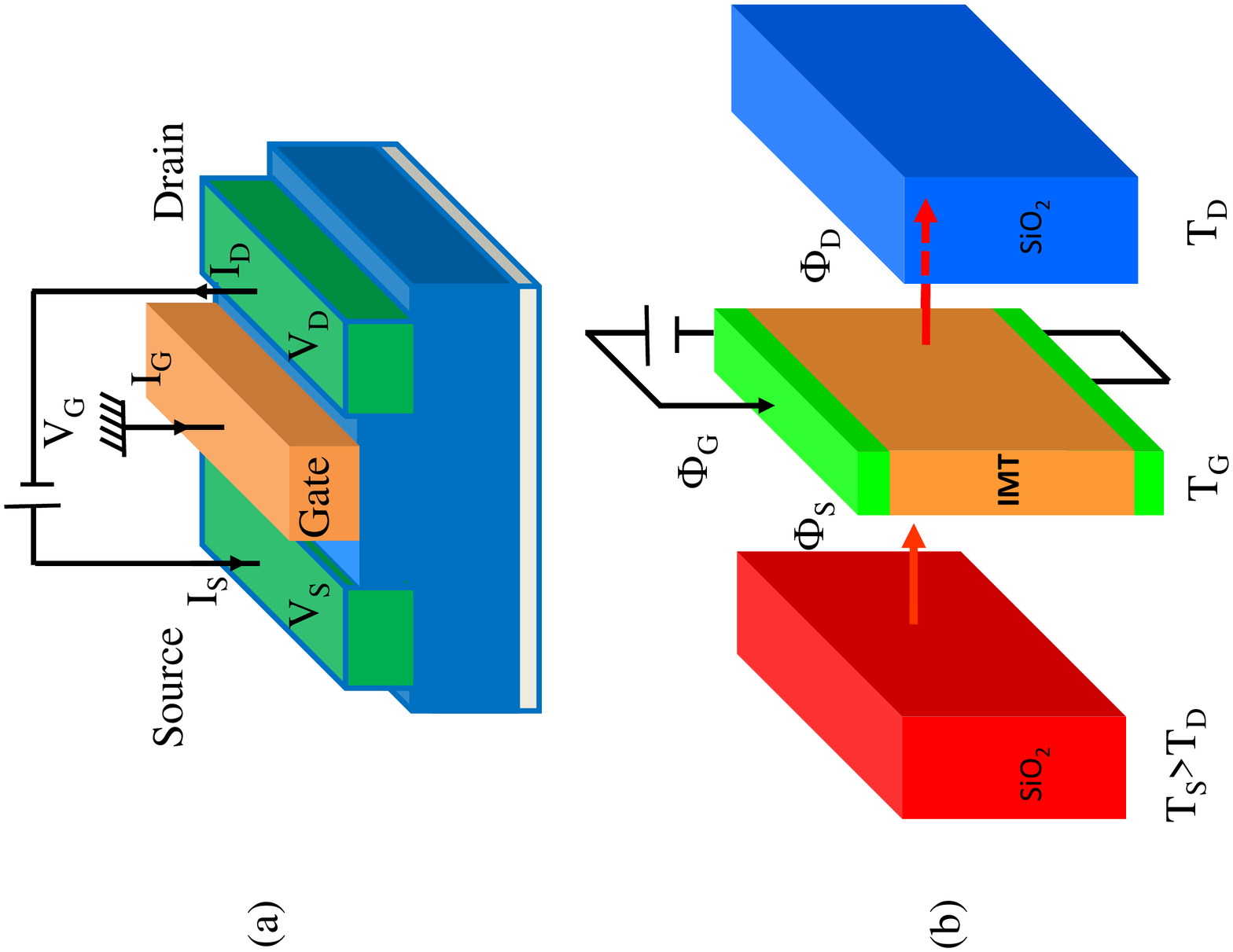} 
\caption{(a) A classical field-effect transistor is a device with three terminals, the source (S), the gate (G), 
         and the drain (D). The gate is used to actively control (by applying a voltage $V$ bias on it) the apparent 
         electric conductivity of the channel between the source and the drain. (b)
         A radiative thermal transistor. A membrane of an IMT material (VO$_2$) acts as the gate 
         between two silica (SiO$_2$) thermal reservoirs (source and drain). The temperature ($T_\rG$)
         of the gate is in steady state between the temperatures of the source and drain 
         ($T_\rS$ and $T_\rD$). $\Phi_\rD$ and $\Phi_\rS$ are the radiative heat fluxes received by the drain and emitted 
         by the source, respectively.}.
\label{transistor}
\end{figure}

The thermal diode allows the heat flow to propagate preferentially in one direction, but such a device does 
not allow for switching, modulating or even amplifying the flux exchanged between a hot and a cold body. However, in 
recent years, physicists have begun to design and test thermal transistors with some success~\cite{BaowenLiEtAl2012}. 
In this paragraph we discuss the possibility to design a radiative thermal analog of electronic transistors as 
depicted in Fig.~\ref{transistor}(a). The classical field-effect transistor (FET) which is composed by three 
basic elements, the drain, the source, and the gate, is basically used to control the flux of electrons 
(the current) exchanged in the channel between the drain and the source by changing the voltage bias applied 
on the gate. The physical diameter of this channel is fixed, but its effective electrical diameter can be 
varied by the application of a voltage on the gate. A small change in this voltage can cause a large variation 
in the current from the source to the drain. 

Recently, a thermal counterpart of the FET has been proposed~\cite{PBA_PRL2014} to control the near-field radiative 
heat exchanges between two bodies. This near-field thermal transistor (NFTT) is depicted in Fig.~\ref{transistor}(b)
and basically consists of a  source and the drain, labeled by the indices S and D, which are maintained at 
temperatures $T_\rS$ and $T_\rD$ using thermostats where $T_\rS > T_\rD$ so that we have a net heat flux from the source
towards the drain. A thin layer of IMT material labeled by G of width $\delta$ is placed between the source 
and the drain at a distance $d$ from both media and functions as the gate. This IMT material is able, as we 
have seen before, to qualitatively and quantitatively change its optical properties through a small 
change of its temperature around a critical temperature $T_\rc$. As in the previous paragraph we describe 
here the operating modes of radiative transistor using VO$_2$ as such IMT material. The choice of IMT depends 
on the operating temperature of the transistor. If the transistor should operate around $T = 500\,{\rm K}$ then 
VO$_2$ could be replaced by LaCoO$_3$, for instance. For the source and the drain we use again SiO$_2$. 
Hence we have in principle two oppositely biased diodes as discussed in the previous section which are connected 
in series so that the NFTT (for which we had a field-effect transistor in mind) can also be regarded 
as a bipolar transistor.

Without external excitation, the system will reach its steady state for which the net flux $\Phi_\rG$ 
received by the intermediate medium, the gate, is zero by heating or cooling the gate until it reaches 
its steady state or equilibrium temperature $T_\rG^{\rm eq}$. In this case the gate temperature 
$T_\rG$ is set by the temperature of the surrounding media, i.e.\ the drain and the source. When a certain amount 
of heat is added to or removed from the gate for example by applying a voltage difference through a 
couple of electrodes as illustrated in Fig.~\ref{transistor}(b) or by extracting heat using Peltier 
elements, its temperature can be either increased or reduced around its equilibrium temperature $T_G^{eq}$. 
This external action on the gate allows to tailor the heat flux $\Phi_\rS$ between the source and the gate 
and the heat flux $\Phi_\rD$ between the gate and the drain. 

In order to see, if this system can be operated as a thermal transistor we need first to
determine the radiative heat flux in this system, which is a little bit more complex then the system
studied for the radiative thermal diode. In a three-body system the radiative flux received by the drain 
takes the form~\cite{Messina}
\begin{equation}
  \Phi_{\rD}= \int_0^\infty\!\frac{d\omega}{2\pi}\,\phi_\rD(\omega,d), 
\label{Eq:Flux_D}
\end{equation}
where the spectral heat flux is given by
\begin{equation}
\begin{split}
  \phi_{\rD} &= \sum_{j = \{\rm s,p\}}\int\! \frac{{\rm d}^2 \boldsymbol{\kappa}}{(2 \pi)^2} \, 
                \bigl[\Theta_{\rS\rG}(\omega)\mathcal{T}^{\rS/\rG}_j(\omega,\boldsymbol{\kappa}; d)\\
                &\quad+ \Theta_{\rm GD}(\omega)\mathcal{T}^{\rG/\rD}_j(\omega,\boldsymbol{\kappa}; d)\bigr].
\end{split}
\label{Eq:Flux_D_1}
\end{equation}
This time $\mathcal{T}^{\rS/\rG}_j$ and $\mathcal{T}^{\rG/\rD}_j$ denote the transmission coefficients of each mode $(\omega,\boldsymbol{\kappa})$ 
between the source and the gate and between the gate and the drain for both polarization states $j=\rs,\rp$. In the above relation 
$\Theta_{ij}$ denotes the difference of functions $\Theta(\omega,T_i)$ and $\Theta(\omega,T_j)$. 
According to the N-body near-field heat transfer theory presented in Ref.~\cite{Messina}, the transmission coefficients $\mathcal{T}^{\rS/\rG}_j$ and 
$\mathcal{T}^{\rG/\rD}_j$ of the energy carried by each mode written in terms of optical reflection coefficients $\rho_{E,j}$ ($\rE = \rS, \rD, \rG$) 
and transmission coefficients $\tau_{E,j}$ of each basic element of the system and in terms of reflection coefficients 
$\rho_{EF,j}$ ($E = \rS, \rD, \rG$ and $F = \rS, \rD, \rG$) of couples of elementary elements~\cite{Messina}
\begin{equation}
\begin{split}
  &\mathcal{T}^{\rS/\rG}_{j}(\omega,\boldsymbol{\kappa},d)\\
  &=\frac{4\mid\tau_{\rG,j}\mid^2 \Im(\rho_{\rS,j})\Im(\rho_{D,j})\re^{-4\gamma d}}{\mid 1-\rho_{\rS\rG,j}\rho_{\rD,j}\re^{-2\gamma d}\mid^2\mid1-\rho_{\rS,j}\rho_{\rG,j}\re^{-2\gamma d}\mid^2},\\
  &\mathcal{T}^{\rG/\rD}_{j}(\omega,\boldsymbol{\kappa},d)=\frac{4 \Im(\rho_{\rS\rG,j})\Im(\rho_{\rD,j})\re^{-2\gamma d}}{\mid 1-\rho_{\rS\rG,j}\rho_{\rD,j}\re^{-2\gamma d}\mid^2}
\end{split}
\label{Trans}
\end{equation}
introducing the imaginary part of the wavevector normal to the surfaces in the multilayer structure $\gamma = \Im(k_{z0})$. 
Similarly the heat flux from the source towards the gate reads
\begin{equation}
\begin{split}
  \phi_{\rS} &= \sum_{j = \{\rm s,p\}}\int\! \frac{{\rm d}^2 \boldsymbol{\kappa}}{(2 \pi)^2} \, \bigl[\Theta_{\rD\rG}(\omega)\mathcal{T}^{\rD/\rG}_j(\omega,\boldsymbol{\kappa}; d)\\
             &\quad + \Theta_{\rG\rS}(\omega)\mathcal{T}^{\rG/\rS}_j(\omega,\boldsymbol{\kappa}; d)\bigr]
\end{split}
\label{Eq:Flux_S_1}
\end{equation}
where the transmission coefficients are analog to those defined in Eq.~(\ref{Trans}) and can be obtained making the substitution $\rS \leftrightarrow \rD$.
At steady state, the net heat flux received/emitted by the gate which is just given by the heat flux from the source to the gate minus the heat flux from the gate to the drain vanishes, i.e. $\Phi_\rS = \Phi_\rD$ or 
\begin{equation}
 \Phi_\rG = \Phi_\rS - \Phi_\rD = 0.
\label{Eq:equilibrium}
\end{equation}
This relation allows us to identify the gate equilibrium temperature $T^{\rm eq}_\rG$ (which is not necessary unique because of 
the presence of bistability mechanisms~\cite{PBA_PRL2014}) for given temperatures $T_\rS$ and $T_\rD$. Note that out of steady state, 
the heat flux received/emitted by the gate is $\Phi_\rG = \Phi_\rS - \Phi_\rD \neq 0$. If $\Phi_\rG < 0$ ($\Phi_\rG > 0$) an external 
flux is added to (removed from) the gate by heating (cooling). 

To illustrate the different operating modes of the NFTT we consider now our system of a silica source and
a silica drain with a VO$_2$ gate in between. We set $T_\rS=360\, {\rm K}$ and $T_D = 300\, {\rm K}$
and choose a separation distance between the source and the gate and between the gate and the drain 
of $d = 100\, {\rm nm}$. The thickness of the gate layer is set to $\delta = 50\, {\rm nm}$. Therefore,
the NFTT operates in the near-field regime (of course, it would also work in the far-field regime).
The equilibrium temperature of the gate is obtained by solving the transcendental equation (\ref{Eq:equilibrium}). 
Here we find $T^{\rm eq}_G=332\,{\rm K}$ which is close to the critical temperature $T_\rc \approx 340\,{\rm K}$ of VO$_2$ which means
that in the steady-state situation the gate is in its insulating phase. In this situation the surface modes
of the gate and the source and the drain can couple which results in large heat fluxes $\Phi_\rS$ and $\Phi_\rD$.
On the other hand, when the temperature of the gate is increased by external heating to values larger than $T_\rc$
then VO$_2$ undergoes a phase transition towards its metallic phase. In this phase we can expect from the
discussion of the diode that the coupling between the gate and the source and the drain will be much less efficient
so that  $\Phi_\rS$ and $\Phi_\rD$ can be expected to drop. In the transition regime around $T_\rc$ we model
the effective permittivity $\epsilon^{\rm eff}_{\parallel/\perp}$ of VO$_2$ parallel and perpendicular to the optical axis by a simple effective medium theory (EMT) ansatz 
\begin{equation}
  \epsilon^{\rm eff}_{\parallel/\perp} = f \epsilon^{\rm m}_{\parallel/\perp} + (1 - f)\epsilon^{\rm d}_{\parallel/\perp}
\end{equation}
and by a Bruggemann model (BM)
\begin{equation}
   f \frac{\epsilon^{\rm d}_{\parallel/\perp} -  \epsilon^{\rm eff}_{\parallel/\perp}}{\epsilon^{\rm d}_{\parallel/\perp} +  2 \epsilon^{\rm eff}_{\parallel/\perp}}  + (1 - f)  \frac{\epsilon^{\rm m}_{\parallel/\perp} -  \epsilon^{\rm eff}_{\parallel/\perp}}{\epsilon^{\rm m}_{\parallel/\perp} +  2 \epsilon^{\rm eff}_{\parallel/\perp}} = 0
\end{equation}
which is inspired by the measurements and modelling in Ref.~\cite{Mott}. Here $\epsilon^{\rm d}_{\parallel/\perp}$ is the permittivity of VO$_2$ in its insulating phase
and $\epsilon^{\rm m}_{\parallel/\perp}$ is the permittivity of VO$_2$ in its metallic phase. The filling fraction $f$ is chosen to be linear in $T_\rG$ such that the
phase transition starts at $T_\rG = 341.2\,{\rm K}$ and ends at $T_\rG = 345.2\,{\rm K}$. Of course, this model can be improved by
using measurements of the permittivities in the transition region. 

\begin{figure}
\includegraphics[scale=0.3,angle=0]{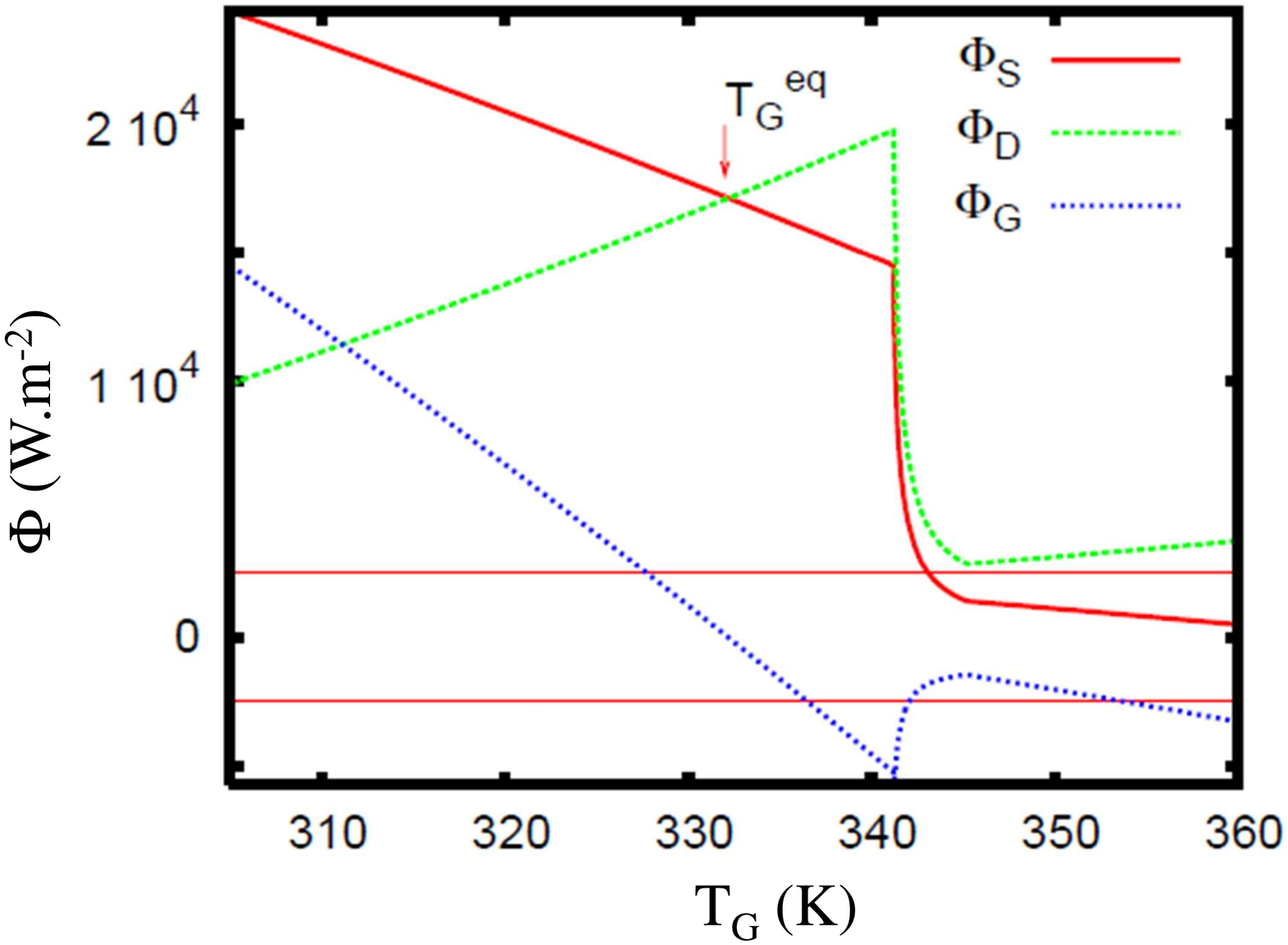} 
\caption{Heat fluxes inside the near-field thermal transistor using a  
         VO$_2$ gate of thickness $\delta= 50\,{\rm nm}$ located at a distance 
         $d = 100\,{\rm nm}$ between two massive silica samples maintained at 
         temperatures $T_\rS = 360\,{\rm K}$ and $T_\rD = 300\,{\rm K}$.
         By changing the flux $\Phi_\rG$ supplied to the gate the temperature $T_\rG$ 
         is changed which allows to operate the NFTT as a 
         thermal switch, a thermal modulator and a thermal amplifier. }.
\label{modes}
\end{figure}

\begin{figure}
  \includegraphics[scale=0.3,angle=0]{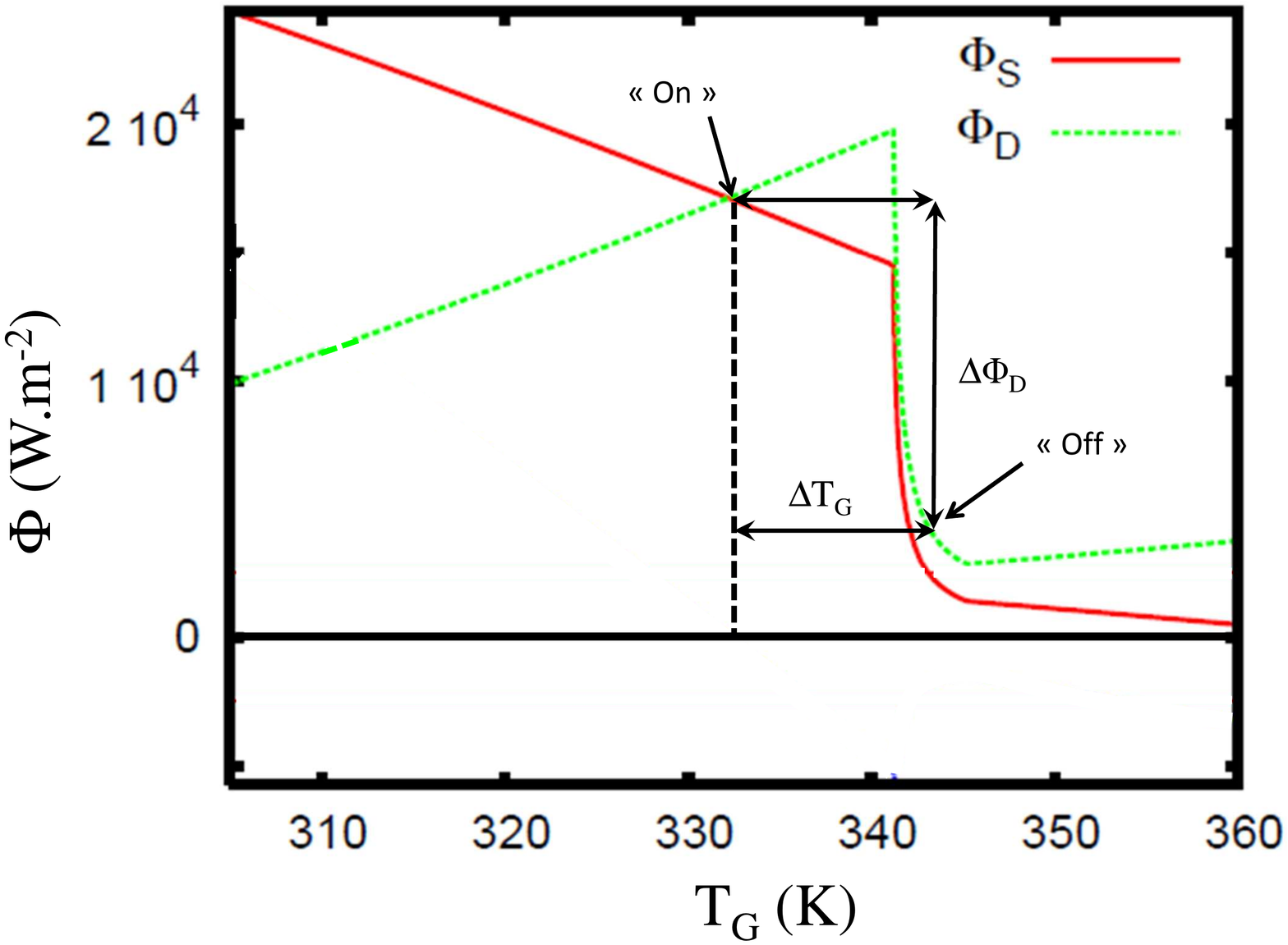} 
     \caption{Switching mode of radiative transistor: When the gate temperature $T_\rG$ is 
              slightly increased above its critical temperature $T_\rc$ the heat 
              flux $\Phi_D$ received by the drain falls from a relatively large value, the so-called "on" mode,
              to a value which is about one order of magnitude smaller, the so-called "off" mode.}
  \label{switch}
  \end{figure}

\begin{figure}
\includegraphics[scale=0.3,angle=0]{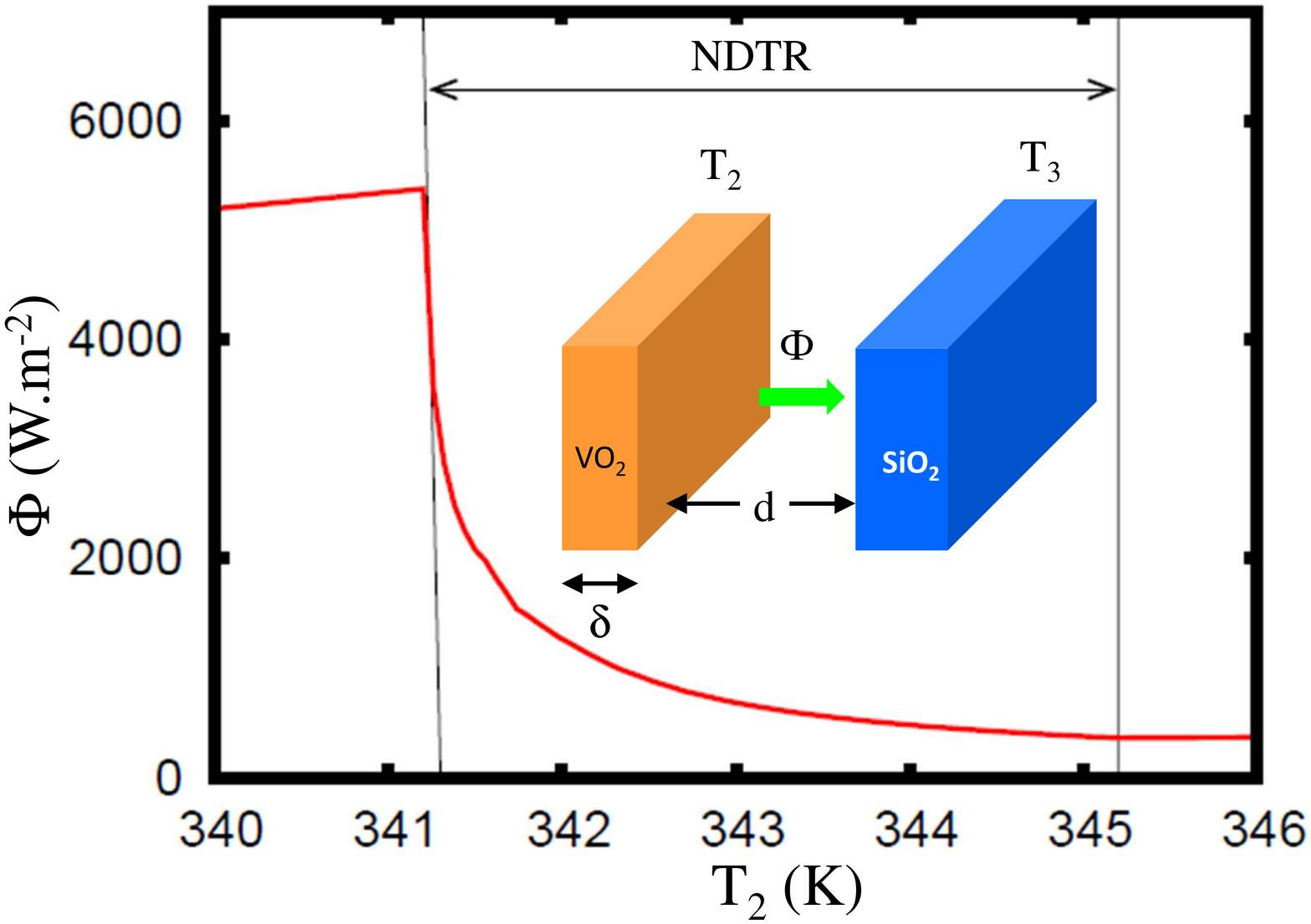} 
\caption{Illustration of the negative differential conductance (NDTC) between a VO$_2$ layer $\delta=50\,{\rm nm}$ thick 
         and a massive SiO$_2$ sample separated by a vacuum gap of $d = 100\,{\rm nm}$. The flux $\Phi$ exchanged between both media 
         is in the phase-transition region $T_2 \in [341.2\,{\rm K},345.2\,{\rm K}]$ a decreasing function of the VO$_2$ 
         temperature $T_2$. Outside this transition region the flux $\Phi$ increases with temperature. Here the temperature of SiO$_2$
         is fixed to $300\,{\rm K}$.}
\label{NDTR}
\end{figure}

Using the above parameters together with the EMT model for the permittivity of VO$_2$ we find
the resulting heat fluxes $\Phi_\rS$, $\Phi_\rD$ and $\Phi_\rG$ as shown in Fig.~\ref{modes}.
These results illustrate that the NFTT can operate as (i) a thermal switch, (ii) a thermal modulator 
or (iii) a thermal amplifier. But let us discuss these functions of the NFTT in more detail: 
\begin{flushleft}
(i)\underline{\textit{Thermal switching :}}
\end{flushleft}
  An increase of $T_\rG$ by about 10 degrees starting from $T_\rG = T_\rG^{\rm eq}$ leads, as clearly shown in 
  Fig.~\ref{switch}, to a reduction of heat flux received by the drain and lost by the source by 
  about one order of magnitude. That means the NFTT can be used in two operating modes where 
  $T_\rG$ is slightly below or above the critical temperature $T_\rc$, where in the case $T_\rG < T_\rc$ 
  we have a large heat flux to the drain which can be assigned as the "on" mode 
  and for $T_\rG > T_\rc$ the heat flux towards the drain drops by one order of 
  magnitude which can be assigned as the "off" mode. The thermal inertia of the gate as well as its phase 
  transition delay defines the timescale at which the switch can operate. Usually the thermal inertia
  limits the speed to some microseconds~\cite{Tschikin,Dyakov2}.
\begin{flushleft}
(ii)\underline{\textit{Thermal modulation:}}
\end{flushleft}
  Over a relatively broad temperature range around $T^{\rm eq}_G$ (strip between the vertical lines in Fig.~\ref{modes}) the heat 
  current $\Phi_\rG$ from or towards the gate remains relatively small compared to the $\Phi_\rS$ and $\Phi_\rG$ (i.e. $\Phi_S\approx\Phi_D$). 
  Within this temperature range the flux received by the drain or lost by the source can be modulated.  A much larger modulation of fluxes 
  can be achieved with the NFTT when setting $T_\rS$ and $T_\rD$ such that $T^{\rm eq}_\rG \approx T_\rc$. Then the flux can be modulated over 
  one order of magnitude by a small temperature change of $T_\rG$. 
\begin{flushleft}
(iii)\underline{\textit{Thermal amplification:}}
\end{flushleft}
  The most important feature for a transistor is its ability to amplify the current or electron flux towards the drain. 
  In the region of phase transition around $T_\rc$ we see in Fig.~\ref{modes} that an increase of $T_\rG$ leads to a drastic 
  reduction of flux received by the drain. As described by Kats et al. in Ref.\cite{Kats} this behavior can be associated (in far-field) to a reduction of the thermal emission. This corresponds to a negative differential thermal conductance (NDTC) as described 
  for SiC in Ref.~\cite{Fan} and illustrated for our configuration in Fig.~\ref{NDTR}. Having a NDTC 
  is the key for having an amplification which is defined as (see for example Ref.~\cite{Casati1})
  \begin{equation}
    \alpha \equiv \biggl|\frac{\partial \Phi_\rD}{\partial \Phi_\rG}\biggl| =  \frac{1}{\biggl|1 - \frac{\Phi_\rS'}{\Phi_\rD'}\biggr|}
  \end{equation}
  where
  \begin{equation}
    \Phi_{\rS/\rD}' \equiv \frac{\partial \Phi_{\rS/\rD}}{\partial T_\rG}.
  \end{equation}

  It can be easily seen from Fig.~\ref{modes} that $\alpha = 1/2$ for $T_\rG$ outside the phase transition region where the material 
  properties of VO$_2$ are more or less independent of $T_\rG$, since $\Phi_S' = - \Phi_D'$ in this region. On the other hand, inside 
  the transition region of VO$_2$ that means for temperatures around $T_\rc$ the material properties of VO$_2$ change drastically 
  showing a NDTC which leads to an amplification of about 4 regardless of using the EMT or BM for modelling the permittivity
  in the transition region~\cite{PBA_PRL2014}.  

\begin{figure}
\includegraphics[scale=0.3,angle=0]{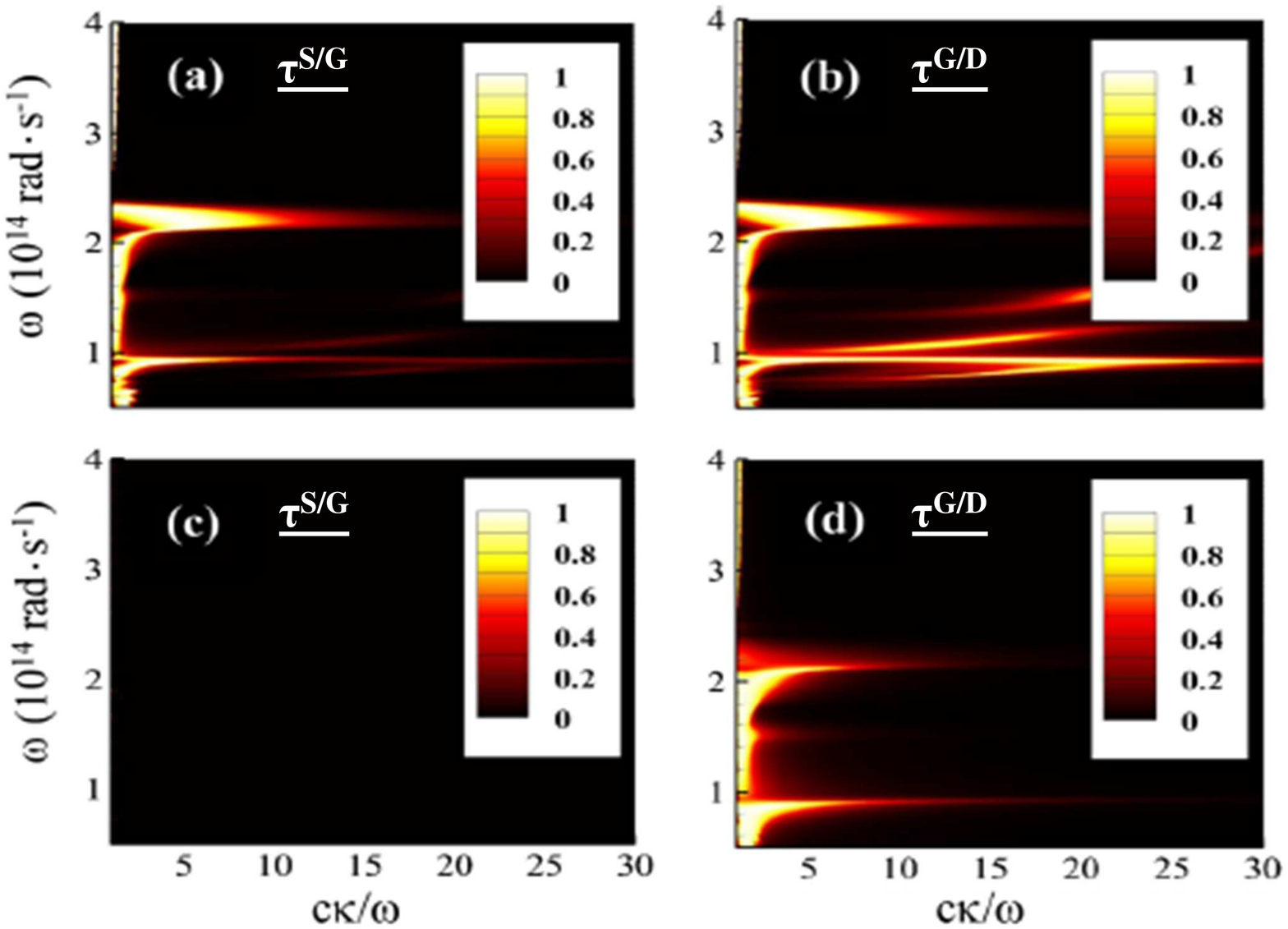} 
\caption{Efficiency of the mode coupling in ($\omega,\boldsymbol{\kappa}$)-space for a SiO$_2$-VO$_2$-SiO$_2$ system 
         ($\delta=50\,{\rm nm}$ and $d=100\,{\rm nm}$). (a) $\mathcal{T}^{\rS/\rG}_\rp$ and (b) $\mathcal{T}^{\rG/\rD}_\rp$ with 
         VO$_2$ in its insulating state. (c) $\mathcal{T}^{\rS/\rG}_\rp$ and (d) $\mathcal{T}^{\rG/\rD}_\rp$ with VO$_2$ in its 
         metallic state. Wien's frequency (where the heat transfer is maximal) at $T=340\,{\rm K}$ is
         $\omega_{\rm Wien}\approx 1.3\times 10^{14}\,{\rm rad/s}$. }
\label{transmission}
\end{figure}

To explain the strong variations of heat flux observed inside our NFTT, let us examine hereafter 
the transmission coefficients (only p polarisation) of the energy carried by the modes $(\omega,\boldsymbol{\kappa})$ 
through such a three-body system which are plotted in Fig.~\ref{transmission}. When the gate is in its insulating 
state, $\mathcal{T}^{\rS/\rG}_\rp$, which represents the exchange between the source and the drain 
mediated by the presence of the gate [see Fig.~\ref{transmission}(a)], and $\mathcal{T}^{\rG/\rD}_\rp$, 
which  corresponds to the exchange between the couple source-gate treated as a unique body and the 
drain [see Fig.~\ref{transmission}(b)] shows an efficient coupling of modes between the different 
blocks of the system around the resonance frequencies $\omega_{\rm SPP1}\approx 1\times 10^{14}\,{\rm rad/s}$ 
and $\omega_{\rm SPP2}\approx 2\times 10^{14}\,{\rm rad/s}$ of surface waves (surfaces phonon-polaritons) 
supported by both the source and the drain. Below $T_\rc$ all parts of the system support surface waves 
in the same frequency range close to the thermal peak frequency $\omega_{\rm Wien}\approx 1.3\times 10^{14}\,{\rm rad/s}$. 
The anti-crossing curves which appear in Fig.~\ref{transmission}(a) and (b) result 
from the strong coupling of silica surface phonon-polaritons (SPPs) and the surface waves 
(symmetric and antisymmetric ones) supported by the thin VO$_2$ layer. Beyond $T_\rc$ the gate becomes 
metallic and it does not support surface waves anymore. In this case $\mathcal{T}^{\rS/\rG}_\rp$ 
[see Fig.~\ref{transmission}(c)] vanishes owing to the field screening by the gate. Moreover, as is clearly 
shown in Fig.~\ref{transmission}(d), the coupling of modes between the source-gate and the drain 
at the frequency of surface waves is less efficient for the large parallel values of $\kappa$ reducing so 
the transfer of heat towards the drain, i.e.\ the number of participating modes decreases~\cite{BiehsEtAl2010,JoulainPBA2010}.

%
%

\section{Volatile radiative memory}
Manipulating radiative heat fluxes is only a first step towards the development of a contactless technology 
for the thermal management of systems at tha nano and macroscale. In this last section, we examine 
the possibility to store information or energy for arbitrary long time using thermal photons by introducing 
the concept of a radiative thermal memory. The concept of thermal memory is closely related to the thermal 
bistability of a system that means to the presence of at least two temperature set's which lead to 
a steady state~\cite{BaowenLi3}. 

\begin{figure}[Hhbt]
\includegraphics[scale=0.3]{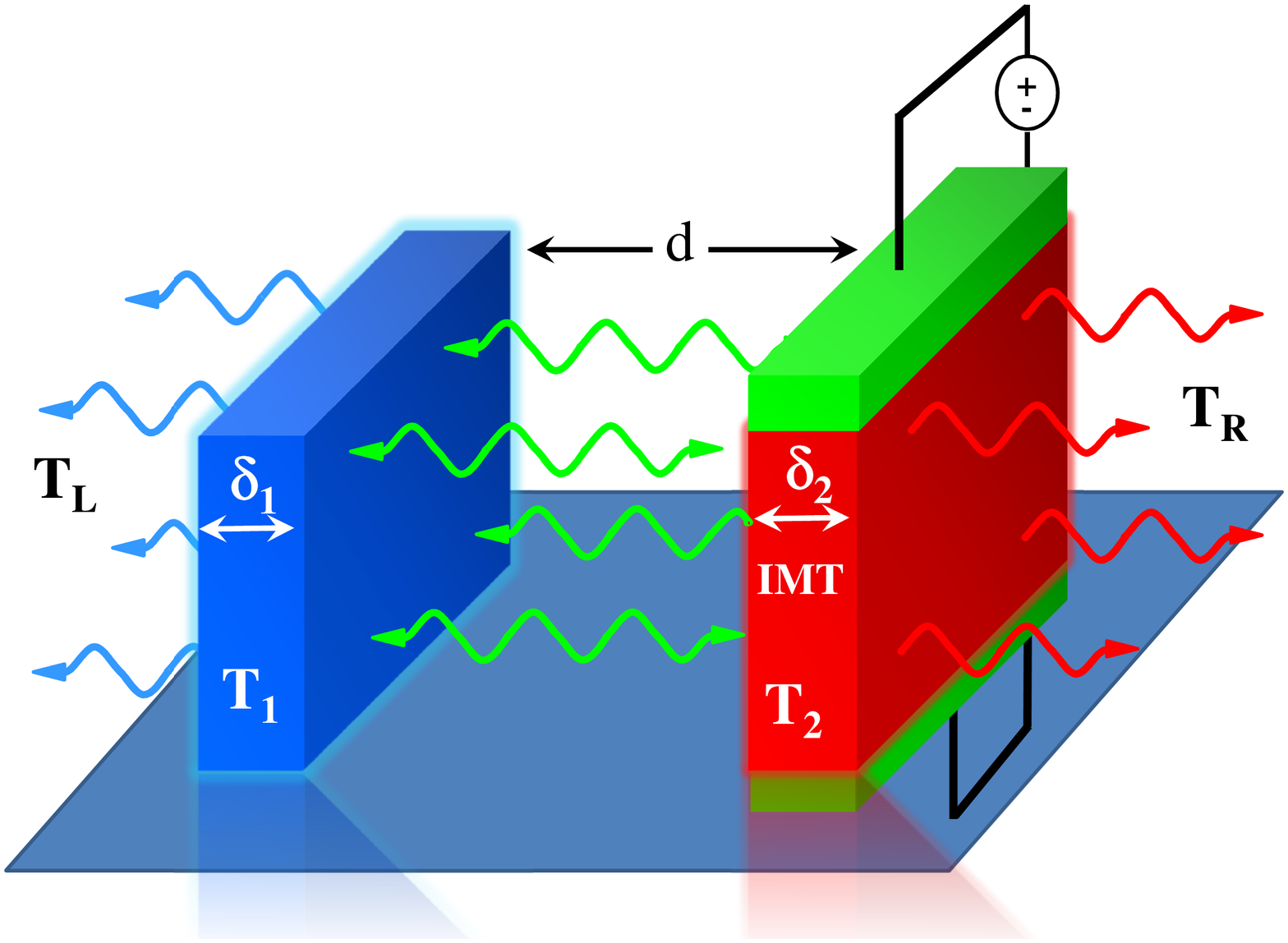}
\caption{Sketch of a radiative thermal memory. A membrane made of an IMT material is placed at a distance 
$d$ from a dielectric layer. The system is surrounded by two thermal baths at different temperatures $T_{\rm L}$ and $T_{\rm R}$. 
The temperature $T_2$ can be increased or reduced either by Joule heating by applying a voltage difference through a couple of 
electrodes or by using Peltier elements.} 
\label{memory}
\end{figure}

To illustrate this, let us consider a concrete system as depicted in Fig.~\ref{memory} composed by two 
parallel homogeneous membranes made of VO$_2$ and SiO$_2$. These slabs have finite thicknesses $\delta_{1}$ and $\delta_{2}$ 
and are separated by a vacuum gap of distance $d$. The left (right) membrane is in contact with a thermal bath 
having a temperature $T_{\rm L}$ ($T_{\rm R}$), where $T_{\rm L} \neq T_{\rm R}$. In a practical point of 
view, the field radiated by these baths can be produced by two external blackbodies. The membranes themselves 
interact on the one hand through the intracavity fields and on the other with the thermal baths which can be 
thought to be produced by external media. In that sense, the system is driven by many-body interactions in contrast
to the diode in section II which is only a two-body system.

The heat flux across any plane $z = \bar{z}$ parallel to the interacting surfaces can again be evaluated
using Rytov's fluctuational electrodynamics~\cite{Rytov,Polder}. For the sake of simplicity we assume that 
the separation distance $d$ is large enough compared to the thermal wavelengths 
[i.e. $d\gg\max(\lambda_{T_i}=c\hbar/(\kb T_{i})$, $i = 1,2,{\rm L}, {\rm R}$ ] so that near-field heat exchanges 
can be neglected (a memory using the near-field effect was proposed quite resently in Ref.~\cite{DyakovMemory}). 
In this case we obtain
\begin{equation}
\begin{split}
     \varphi(\bar{z}) = 2\epsilon_0c^2 \!\!\sum \limits_{\underset{\phi=\left\{ +,-\right\}}{j=s,p}}\int_{0}^{\infty}\!\!\frac{d\omega}{2\pi}\int\!\!\! \frac{{\rm d}^2 \boldsymbol{\kappa}}{(2 \pi)^2}\frac{\phi k_{z0}}{\omega} \mathfrak{C}_j^{\phi,\phi}(\omega,\boldsymbol\kappa),
\label{Eq:base}
\end{split}
\end{equation}
where the field correlators~\cite{Messina3}
\begin{equation}
\mathfrak{C}_j^{\phi,\phi'}(\omega,\boldsymbol{\kappa})= \frac{1}{2} \langle[E_j^\phi(\omega,\boldsymbol{\kappa})E_j^{\phi'\dagger}(\omega,\boldsymbol{\kappa})+E_j^{\phi'\dagger}(\omega,\boldsymbol{\kappa})E_j^\phi(\omega,\boldsymbol{\kappa})]\rangle
\end{equation}
of local field amplitudes in polarization $j$ can be expressed (see the supplemental material in~\cite{Slava} for details)  in terms of 
reflection and transmission operators $\mathfrak{R}_i^{\pm}$ and $\mathfrak{T}_i^{\pm}$  of the layer $i$  toward the right ($+$) and the left ($-$), as
\begin{equation}
\begin{split}
  \mathfrak{C}_1^{+,+}=\mathcal{S}(T_1)(1- \mid \mathfrak{R}_1^{+} \mid^2 -\mid \mathfrak{T}_1^{+} \mid^2 ),\\
  \mathfrak{C}_2^{-,-}=\mathcal{S}(T_2)(1- \mid \mathfrak{R}_2^{-} \mid^2 -\mid \mathfrak{T}_2^{-} \mid^2 ),\\
  \mathfrak{C}_L^{+,+}=\mathcal{S}(T_L),\;\;\;\;\;\;\;\;\;\;\;\;\;\;\;\;\;\;\;\;\;\;\;\;\;\;\;\;\;\;\;\;\;\;\\
  \mathfrak{C}_R^{-,-}=\mathcal{S}(T_R),\;\;\;\;\;\;\;\;\;\;\;\;\;\;\;\;\;\;\;\;\;\;\;\;\;\;\;\;\;\;\;\;\;\;\\
\end{split}\label{Eq:cavity5}
\end{equation}
with 
\begin{equation}
  \mathcal{S}(T)=\pi\frac{\omega}{\epsilon_0 c^2}\Theta(\omega,T)\Re\biggl(\frac{1}{k_z}\biggr).
\end{equation}
Here $k_z$ denotes the normal component of wave vector in the medium of consideration. Using expression (\ref{Eq:base}) we can calculate the
net flux $\Phi_1=\varphi(0)-\varphi(-\delta_1)$ [resp. $\Phi_2=\varphi(d+\delta_2)-\varphi(d)$] received by the first (resp. second) membrane.

\begin{figure}[Hhbt]
\includegraphics[scale=0.3]{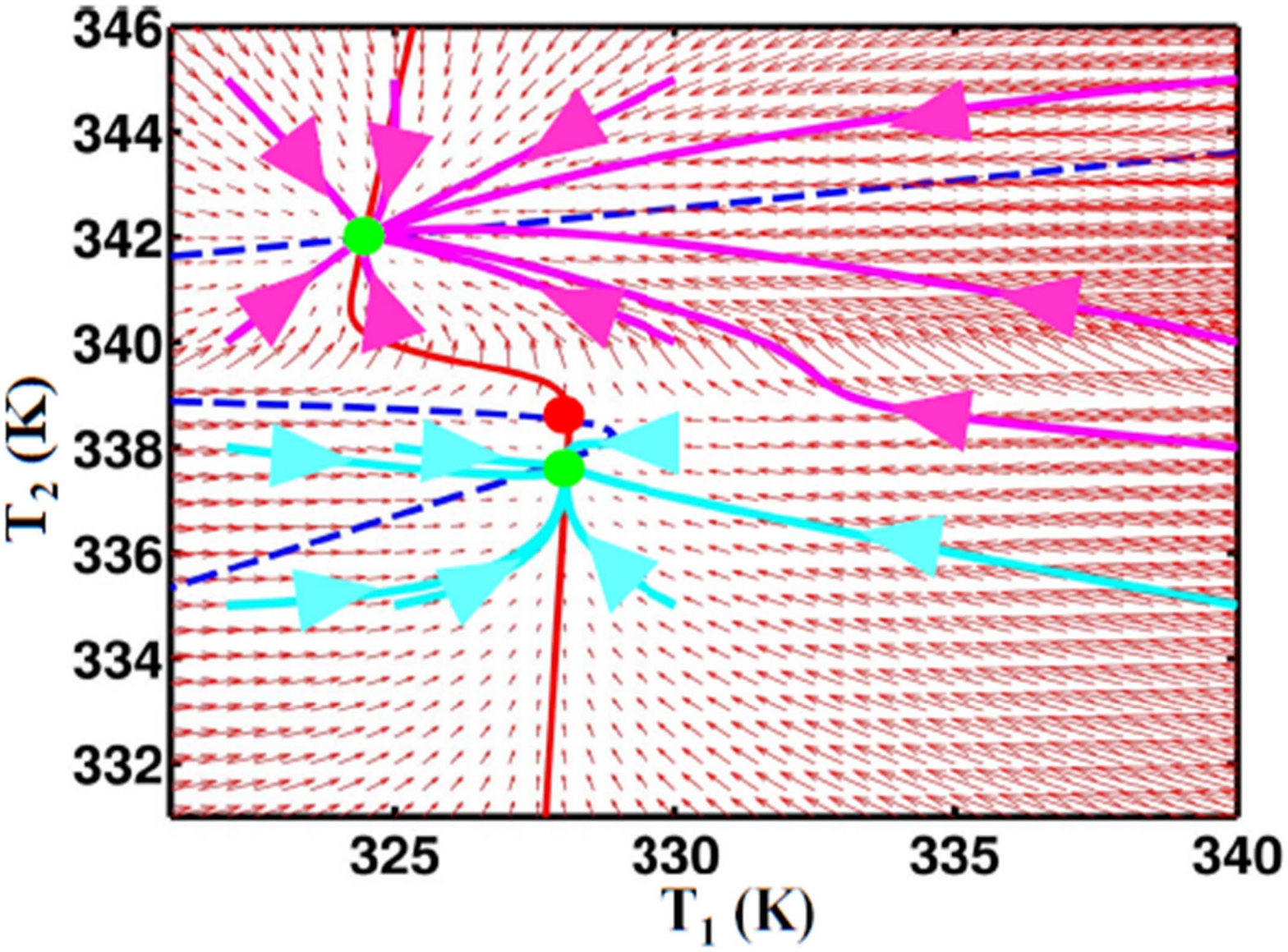}
\caption{Trajectories of temperatures (i.e. phase portrait) for different initial conditions in the plane ($T_1,T_2$) in a two membrane SiO$_2$/VO$_2$ system with $\delta_1=\delta_2=1\,\mu{\rm m}$. The  blue dashed and red solid lines represent the local equilibrium conditions $\Phi_1 = 0$ and $\Phi_2 = 0$ of each membrane. The green (red) points denote the stable (unstable) 
global steady-state temperatures, $(T_1^{(1)},T_2^{(1)}) = (328.03\,{\rm K},337.77\,{\rm K})$, $(T_1^{(2)},T_2^{(2)}) = (328.06\,{\rm K},338.51\,{\rm K})$, and  $(T_1^{(3)},T_2^{(3)}) = (324.45\,{\rm K},341.97\,{\rm K})$. The red arrows represents the vector field $\boldsymbol{\Phi}$. The temperature of thermal reservoirs are $T_{\rm L}=320\,{\rm K}$ and $T_{\rm R}=358\,{\rm K}$. } 
\label{bistability}
\end{figure}

The time evolution of temperatures $T_1$ and $T_2$ of the two membranes are solution of the following nonlinear 
coupled system of differential equations
\begin{equation}
  \partial_t \mathbf{T} = \boldsymbol{\Phi} + \mathbf{Q}
  \label{Eq:diff}
\end{equation}
where we have introduced the vectors $\mathbf{T} \equiv \bigl(T_1(t),T_2(t)\bigr)^\rt$,
 $\boldsymbol{\Phi} \equiv \bigl(\Phi_1(T_1,T_2)/I_1, \Phi_2(T_1,T_2)/I_2\bigr)^\rt$,
and $\mathbf{Q} \equiv (Q_1\delta_1/I_1,Q_2\delta_2/I_2)^\rt$. Here  $Q_i$ ($i = 1,2$) is the power per 
unit volume which can be added to or extracted from both membranes by applying a voltage difference 
through a couple of electrodes as illustrated in Fig.~\ref{memory} or by using Peltier elements.
Furthermore, we have introduced the thermal inertia of both membranes as $I_i \equiv C_i \rho_i \delta_i$, 
where $C_i$ and $\rho_i$ are the heat capacity and the mass density of each material. By writing 
down this set of equations we have neglected any temperature variation inside the membranes 
which is a very good approximation given that the conductivity inside the membranes is much larger than between 
the membranes. When assuming that no energy is directly added to or removed from the membranes, 
then $\mathbf{Q} = \mathbf{0}$. In this case, the steady-state solution is given 
by $\boldsymbol{\Phi} = \mathbf{0}$. Hence $\Phi_1$ and $\Phi_2$ vanish for the same couple of
temperatures $(T_1^{({\rm st-st})},T_2^{(\rm st-st)})$. Considering for an instant membrane 2 only, 
then the existence of two equilibrium temperatures where the net flux vanishes ($\Phi_2 = 0$) implies 
that $\Phi_2$ must have a maximum or a minimum between these two temperatures.
Hence, this requires a  negative differential conductive behavior for this membrane which was shown 
in the previous section for VO$_2$. For the whole system it is therefore a precondition to have 
at least one membrane which exihibits negative differential conductive behavior in order to have 
two couples $(T_1^{({\rm st-st})},T_2^{(\rm st-st)})$ of steady-state temperatures.

In Fig.~\ref{bistability}, we show the time evolution of the SiO$_2$-VO$_2$ system without external 
excitation (i.e. $\mathbf{Q} = \mathbf{0}$) when the thermal inertia $I_i$ of both membranes are 
comparable (i.e. $\delta_1\thickapprox\delta_2$) or very different (i.e. $\delta_1\gg\delta_2$). 
The trajectories (the thick pink and turquois lines) are obtained by solving Eq.~(\ref{Eq:diff}) using 
a Runge-Kutta method with adaptative time steps choosing different initial conditions. In this figure, 
the dashed blue (solid red) line represents the local equilibrium temperatures for the first (second) membrane 
that is the set of temperatures couples ($T_1,T_2$) which satisfy the condition $\Phi_1(T_1,T_2)=0$ [$\Phi_2(T_1,T_2)=0$]. 
The intersection of these two lines define the global steady-state temperatures of the system where $\boldsymbol{\Phi} = \mathbf{0}$. 
Only two of three equilibrium points $\mathbf{T}\equiv (T^{eq}_1,T^{eq}_2)^t$ that appear in 
Fig.~\ref{bistability} are stable. These points can be identified by applying a small 
perturbation $\delta\mathbf{T} \equiv (\delta T_1(t),\delta T_2(t))^t$ on them and look at the 
relaxation process.  Then, the time evolution of each perturbation is driven by the following equation
\begin{equation}
  \partial_t (\delta\mathbf{T}) = \mathbf{J} \cdot \delta\mathbf{T},
  \label{Eq:diff}
\end{equation}
where
\begin{equation}
  \mathbf{J} = \left(\begin{array}{cc}
               \frac{1}{I_1} \frac{\partial\Phi_1}{\partial T_1} & \frac{1}{I_1} \frac{\partial\Phi_1}{\partial T_2} \\
               \frac{1}{I_2} \frac{\partial\Phi_2}{\partial T_1} & \frac{1}{I_2} \frac{\partial\Phi_2}{\partial T_2} \end{array}\right)
               \label{Eq:A0}
\end{equation}
is the Jacobian matrix associated to the dynamical system. Its solution reads
\begin{equation}
  \delta\mathbf{ T}(t) = \delta \mathbf{ T}(0)\exp(\mathbf{J}t).
\end{equation} 
Hence, the eigenvalues $\lambda_i$ of $\mathbf{J}$  allow to determine the stability of thermal state. 
In a two membrane SiO$_2$-VO$_2$ system with $\delta_1=\delta_2=1\,\mu{\rm m}$ and with reservoirs 
temperatures $T_{\rm L}=320\,{\rm K}$ and $T_{\rm R}=358\,{\rm K}$ the stability analysis leads to
\begin{center}
\begin{tabular}{|l|c|r|c|c|}
  \hline
  $T^{\rm eq}_1 (K)$ & $ T^{\rm eq}_2 (K)$ & $\lambda_1$ & $\lambda_2$&stability\\
  \hline
 328.03 & 337.77 & -2.58 & -1.37&stable\\
 328.06 & 338.51 & -2.51 &  3.69&unstable\\
 324.45 & 341.97 & -2.31 &  -4.70&stable\\
  \hline
\end{tabular}
\end{center}
where we have used the volumetric densities and the heat capacities 
$\rho_1 = 2180 \, {\rm kg} {\rm m}^{-3}$, $\rho_2 = 4339 {\rm kg} {\rm m}^{-3}$, 
$C_1 = 750 {\rm J} {\rm kg}^{-1} {\rm K}^{-1}$ and $C_2 = 690 {\rm J} {\rm kg}^{-1} {\rm K}^{-1}$. 


\begin{figure}[Hhbt]
\includegraphics[scale=0.35,angle=-90]{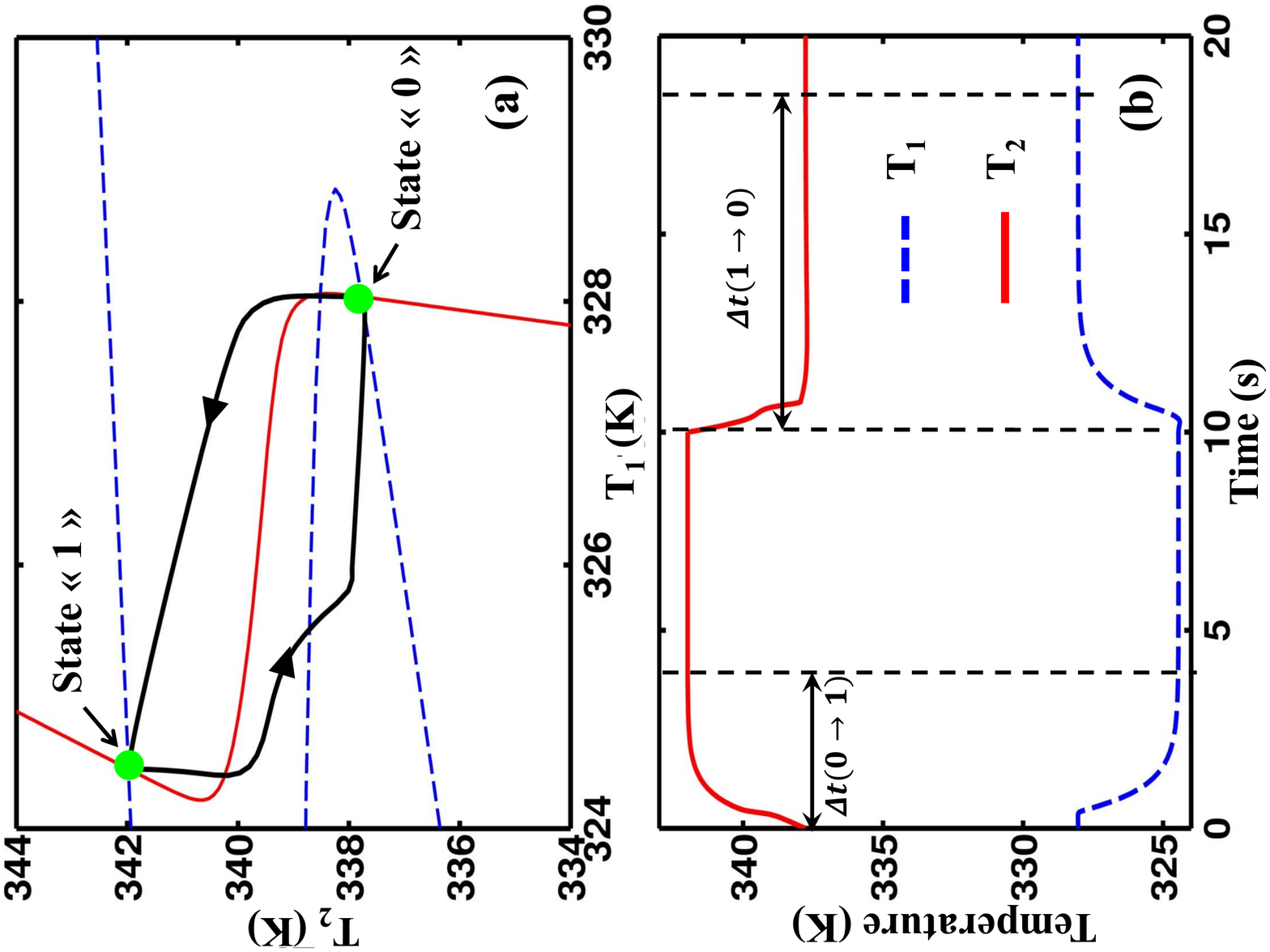}
\caption{(a) Hystereis of the VO$_2$ membrane temperature during a transition between the thermal states "0" and  "1" inside a two membrane SiO$_2$/VO$_2$ system 
with $\delta_1=\delta_2=1\,\mu{\rm m}$. The volumic powers supplied and extracted from the VO$_2$ layer during a time interval $\Delta t_1=0.4\,{\rm s}$ and $\Delta t_2=1.5\,{\rm s}$  
are $Q_2=10^{-2}\,{\rm W} {\rm mm}^{-3}$ and $Q_2=-2.5\times10^{-2}{\rm W}{\rm mm}^{-3}$, respectively. The writing time of state "1" ("0") from the state "0" ("1") is  
$\Delta t(0\rightarrow 1)=4\,{\rm s}$ ($\Delta t(1\rightarrow 0)=8\,{\rm s}$). (b) Time evolution of temperature $T_1(t)$ and $T_2(t)$ of SiO$_2$ and VO$_2$ membranes. 
The thermal states "0" and "1" can be maintained for arbitrary long time provided that the thermostats ($T_L=320\,{\rm K}$ and $T_R=358\,{\rm K}$) remain switched on.} 
\label{memory2}
\end{figure} 

From this result, it becomes apparent that it is possible to use the SiO$_2$-VO$_2$ system as
 a thermal memory~\cite{Slava}. Indeed, the two stable equilibrium temperatures can be 
viewed as two thermal states "0" and  "1" of the system. To switch from one thermal 
state to the other, we need to add or extract power from the system. In the following 
we describe this writing-reading procedure. To this end, we consider the SiO$_2$-VO$_2$ 
system made with membranes of equal thicknesses $\delta_1 = \delta_2=1\,\mu{\rm m}$ which are
coupled to two reservoirs of temperatures $T_\rL=320\,{\rm K}$ and $T_\rR=358\,{\rm K}$. 
Let us define "0" as the thermal state at the temperature $T_2=\min(T^{(1)}_2,T^{(3)}_2)$. 
To make the transition towards the thermal state "1" the VO$_2$ membrane must be heated. 

Step 1 (transition from the state "0" to the state "1"): A volumic power $Q_2=10^{-2}\,{\rm W}{\rm mm}^{-3}$ 
is added to this membrane during a time interval $\Delta t_1\thickapprox 0.4\,{\rm s}$ to reach a region in the plane ($T_1,T_2$) [see  Fig.~\ref{memory2}(a)]
where all trajectories converge naturally (i.e.\ for $Q_2=0$) after some time toward the state "1", the overall transition time 
is $\Delta t(0\rightarrow 1)=4 s$  Fig.~\ref{memory2}(b)].

Step 2 (maintaining the stored thermal information): Since the state "1" is a fixed point, the thermal 
data can be maintained for arbitrary long time provided that the thermal reservoirs are switched on. 
This corresponds basically to the concept of volatile memory in electronics.  

Step 3 (transition from the state "1" to the state "0"): Finally, a volumic power $Q_2=-2.5\times 10^{-2}\,{\rm W}{\rm mm}^{-3}$ 
is extracted from the VO$_2$ membrane during a time interval $\Delta t_2\thickapprox 1.5\,{\rm s}$ to reach a region 
[below $T_2=338\,{\rm K}$ in Fig.~\ref{memory2}(a)] of natural convergence to the state "0" . In this case the transition 
time becomes $\Delta t(1\rightarrow 0)=8\,{\rm s}$. Compared with its heating, the cooling of VO$_2$ does not follow the 
same trajectory [see Fig.~\ref{memory2}(a)] outlining the hysteresis of the system which accompanies its bistable behavior. 
To read out the thermal state of the system a classical electronic thermometer based on the thermo 
dependance of the electric resistivity of membranes can be used.

%
%

\section{Conclusion}
In this paper we have summarized the very recent advances made in optics to manipulate and store the thermal energy with 
contactless devices. We have demonstrated the feasability for contactless thermal analogs of diodes, transistors 
and volatile memories. These devices allow for contactless management of heat flows at macroscale and at subwavelength scales. 
These results pave the way for a novel technology of thermal management. It also suggests the possibility to develop 
contactless thermal analogs of electronic devices such as thermal logic gates for processing information by utilizing 
thermal photons rather than electrons. They also could find broad applications in MEMS/NEMS technologies, to generate 
mechanical work by using microresonators coupled to a transistor as well as in energy storage technology to store 
and release heat upon request.

%
%

\begin{acknowledgments}
S.-A.\ B. and P.\ B.-A acknowledge financial support by the DAAD and Partenariat Hubert Curien Procope Program (project 55923991).
P.\ B.-A acknowledges financial support from the CNRS Energy program.
\end{acknowledgments}

\end{document}